\documentclass[preprint,12pt]{elsarticle}

\usepackage{amssymb}

\usepackage{amsmath,amsfonts}
\usepackage{algorithmic}
\usepackage{algorithm}
\usepackage{array}
\usepackage{multicol} 
\usepackage{arydshln}
\usepackage{graphicx}
\usepackage{amsmath}
\usepackage{amssymb}
\usepackage{booktabs}
\usepackage{times}
\usepackage{epsfig}
\usepackage{graphicx}
\usepackage{amsmath}
\usepackage{amssymb}
\usepackage{booktabs}
\usepackage{multirow}
\usepackage{tabularx}
\usepackage{amssymb}
\usepackage{bbding}
\usepackage{color}
\usepackage{subfigure}
\usepackage{bm}
\usepackage{pifont}
\newcommand{\xmark}{\text{\ding{55}}}
\usepackage[caption=false,font=normalsize,labelfont=sf,textfont=sf]{subfig}
\usepackage{textcomp}
\usepackage{stfloats}
\usepackage{url}
\usepackage{verbatim}
\usepackage{graphicx}

\usepackage{caption}

\journal{Information Sciences}

\begin{document}

\begin{frontmatter}



\title{CFNet: Conditional Filter Learning with Dynamic Noise Estimation for Real Image Denoising}


\author[label1]{Yifan Zuo}
\author[label1]{Jiacheng Xie}
\author[label1]{Yuming Fang}
\author[label2]{Yan Huang}
\author[label1]{Wenhui Jiang}

\address[label1]{organization={School of Information Management, Jiangxi University of Finance and Economics},
             addressline={No. 665 Yuping West Street},
             city={Nanchang},
             postcode={330031},
             state={Jiangxi},
             country={China}} 
             
\address[label2]{organization={Institute of Automation, Chinese Academy of Sciences},
             addressline={No. 95 Zhongguancun East Road},
             city={Beijing},
             postcode={100190},
             state={Beijing},
             country={China}}






\begin{abstract}
A mainstream type of the state of the arts (SOTAs) based on convolutional neural network (CNN) for real image denoising contains two sub-problems, \textit{i.e.,} noise estimation and non-blind denoising. This paper considers real noise approximated by heteroscedastic Gaussian/Poisson Gaussian distributions with in-camera signal processing pipelines. The related works always exploit the estimated noise prior via channel-wise concatenation followed by a convolutional layer with spatially sharing kernels. Due to the variable modes of noise strength and frequency details of all feature positions, this design cannot adaptively tune the corresponding denoising patterns. To address this problem, we propose a novel conditional filter in which the optimal kernels for different feature positions can be adaptively inferred by local features from the image and the noise map. Also, we bring the thought that alternatively performs noise estimation and non-blind denoising into CNN structure, which continuously updates noise prior to guide the iterative feature denoising. In addition, according to the property of heteroscedastic Gaussian distribution, a novel affine transform block is designed to predict the stationary noise component and the signal-dependent noise component. Compared with SOTAs, extensive experiments are conducted on five synthetic datasets and three real datasets, which shows the improvement of the proposed CFNet.
\end{abstract}


\end{frontmatter}


\section{Introduction}
Given noisy observations $\mathbf{y}$, the goal of image denoising is to recover the corresponding clean images $\mathbf{x}$, which has been studied for a few decades. Although additive noise model is assumed that $\mathbf{y}=\mathbf{x}+\mathbf{n}$, where $\mathbf{n}$ represents the noise, it is non-trivial to define the noise $\mathbf{n}$ due to the highly-complicated generation process. In the early research, since the synthetic assumption on noise deviates from real-world situations, the noticeable performance degradation of these models can be observed for real noise removal. 

\begin{figure*}[!t]  
	\centering
	\subfigure[Noisy]{\centering
		\includegraphics[width=0.32\columnwidth]{
		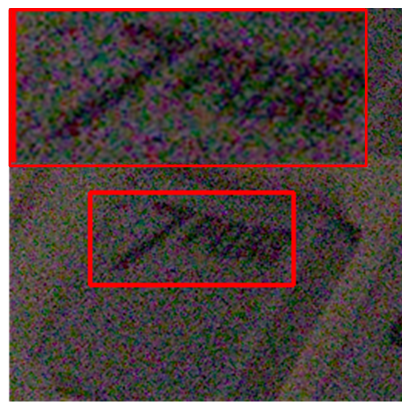}
	}\hspace{-0.06in}
	\subfigure[RIDNet~\cite{ridnet}]{\centering
		\includegraphics[width=0.32\columnwidth]{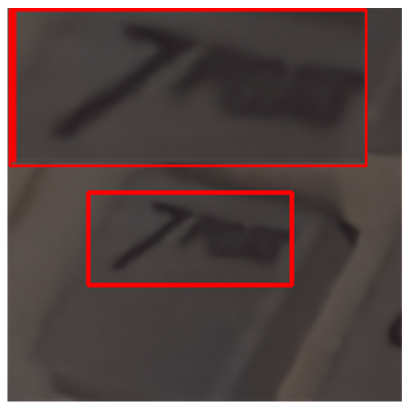}
	}\hspace{-0.06in}	
	\subfigure[VDN~\cite{vdn}]{\centering
		\includegraphics[width=0.32\columnwidth]{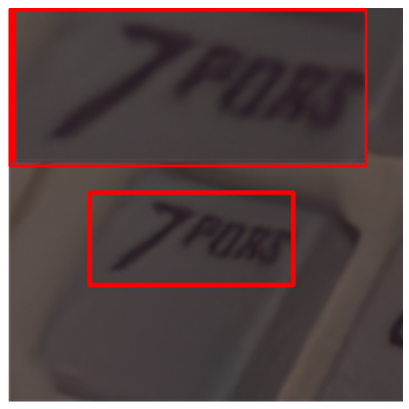}
	}
	\subfigure[InvDN~\cite{invdn}]{\centering
		\includegraphics[width=0.32\columnwidth]{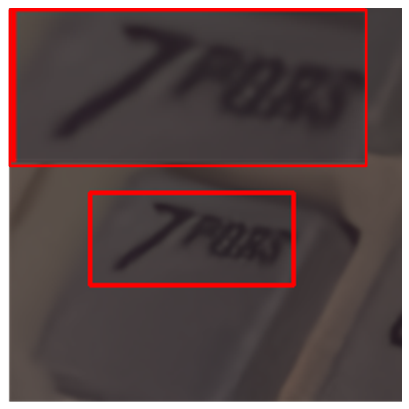}
	}\hspace{-0.06in}	
	\subfigure[\bf{Ours}]{\centering
		\includegraphics[width=0.32\columnwidth]{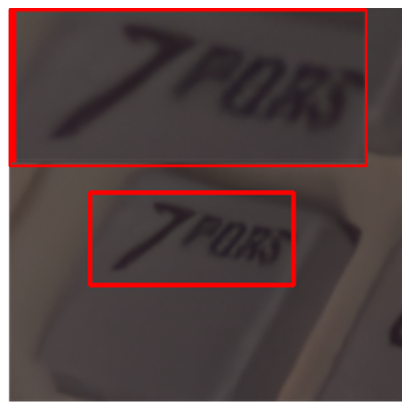}
	}\hspace{-0.06in}
	\subfigure[GT]{\centering
		\includegraphics[width=0.32\columnwidth]{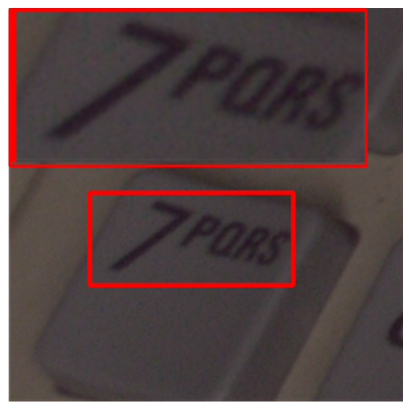}
	}
	\caption{An example of real noise removal from SIDD dataset~\cite{sidd}. (a) The noisy image is recovered by (b) RIDNet~\cite{ridnet}, (c) VDN~\cite{vdn}, (d) InvDN~\cite{invdn} and (e) Ours. Compared with (f) ground truth (GT), our model maintains more details without over-smooth artifact.}
	\label{fig1}
\end{figure*}

Due to strong learning ability of convolutional neural networks (CNNs), some recent works focus on challenging real noise removal~\cite{guo2019toward,aindnet,zamirmirnet}. A mainstream of the state of the arts (SOTAs) always splits real image denoising into two subpropblems, \textit{i.e.,} noise estimation and non-blind denoising. In addition, synthetic heteroscedastic Gaussian/Poisson Gaussian distributions followed by an in-camera signal processing (ISP) pipeline is widely used to approximate real noise, which provides supervision for noise estimation. Thus, such SOTAs use both synthetic noisy images and real-world noisy images to train denoising models. Although significant progress is achieved by CNN-based methods, there are some drawbacks which can be summarized in three aspects. First, for real noisy images, the local windows centered by all pixel positions and feature positions always have different frequency components and suffer from various noise strength. However, the estimated noise maps are usually introduced in SOTAs via channel-wise concatenation followed by a convolutional layer with spatially sharing kernels. Therefore, they cannot adaptively tune the denoising patterns for all feature positions. Second, since most CNN-based SOTAs conduct noise estimation and non-blind denoising in sequential stages, the noise prior cannot be accordingly updated as iterative non-blind feature denoising. Third, existing works always estimate the noise prior of heteroscedastic Gaussian distribution by stacking a few convolutional layers, which is uninterpretable.

To improve interpretability of CNNs, this paper provides more insights based on traditional methods. Accordingly, traditional methods reveal that the optimal denoising pattern should be related to noise strength and frequency details, \textit{i.e.,} edge-preserving denoising. In this paper, a novel CNN-based conditional filter is proposed in which convolutional kernels for different feature positions can be adaptively inferred by the analysis upon noise prior and image features. Furthermore, we bring the thought that alternatively performs noise estimation and non-blind denoising into CNN structure, which continuously updates the noise prior to guide the from-coarse-to-fine feature denoising. In addition, this paper designs iterative affine transform blocks to estimate the translation and the scaling for stationary noise component and signal-dependent noise component of heteroscedastic Gaussian distribution, respectively. Fig.~\ref{fig1} shows an example denoised by the proposed CFNet which recovers more details than three SOTAs. To summary, the main contributions of this paper are listed as below.
\begin{itemize}
\item To the best of our knowledge, this work is the first to propose a conditional filter in which denoising kernels of all feature positions can be adaptively inferred. Therefore, the denoising pattern can be adaptively tuned for local regions by the analysis of image features and noise prior. 
\item Compared with static noise estimation, this paper is the first to propose dynamic noise estimation scheme in CNN structure which alternatively performs noise estimation and non-blind denoising. That is, the noise prior can be accordingly updated to progressively guide feature denoising.
\item A novel noise estimation module is designed by iterative affine transform blocks. Thus, the corresponding translation and scaling can efficiently predict heteroscedastic Gaussian noise distribution. It provides more reasonable model bias than stacking convolutional layers.
\item Compared with SOTAs, the comprehensive and extensive experiments are conducted on five synthetic datasets, \textit{i.e.,} Set12~\cite{dncnn}, BSD68~\cite{bsd68}, CBSD68~\cite{bsd68}, Kodak24~\cite{kodak} and McMaster~\cite{mcmater}, and three real datasets, \textit{i.e.,} SIDD~\cite{sidd}, DND~\cite{dnd} and RN15~\cite{rni}. It shows improvement of the proposed CFNet for synthetic noise removal and real noise removal.
\end{itemize}

The remainder paper is organized as follows. Section II reviews the related works. The proposed CFNet is introduced in Section III. Section IV presents extensive performance comparison to SOTAs. Section V provides in-depth discussion of the proposed CFNet. Finally, Section VI concludes this paper.

\section{Related Works}
In this section, we summarize and discuss the development and recent trends in image denoising. According to assumed noise models, the review is classified into synthetic noise removal and real noise removal. 

\subsection{Synthetic Noise Removal}
In the early research, the synthetic additive white Gaussian noise (AWGN) is first studied for image denoising. The filter-based works remove noise via local filtering which independently refines each pixel position, \textit{e.g.,} neighborhood filters and Non-local means. The representative work~\cite{bm3d} proposes collaborative filtering for grouping similar image blocks. Overall, filter-based methods are efficient at the cost of limited performance by only considering local prior.

By introducing global prior on whole images, the noise is removed by optimization-based methods under the framework of maximum posterior estimation. The handcraft prior is widely expressed by nonlocal self-similarity, sparsity and total variation priors. The representative work~\cite{wnnm} exploits image nonlocal self-similarity based on the weighted nuclear norm minimization. Compared with filter-based counterparts, optimization-based methods show improvement by introducing more complicated global priors.

In addition to hand-craft modeling, learning-based methods provide a data-driven way for image denoising, which presents superior performance for the problems with specific distributions. Based on sparse coding, Aharon \textit{et al.}~\cite{ksvd} propose the K-SVD algorithm to achieve sparse signal representations via adapting dictionaries. Based on low rank prior, Deng \textit{et al.}~\cite{deng2020estimation} model image denoising as a soft-thresholding operation on the singular values of rank-order similar matrices which are built by stacking non-local similar patch vectors. These learning methods are based on shallow models which limits the power of feature representation.

Recently, the ene-to-end training fashion introduced by deep learning reaches significant progress in image denoising. Zhang \textit{et al.}~\cite{dncnn} propose a pioneer CNN-based work using residual learning and batch normalization. Zhang \textit{et al.}~\cite{ffdnet} present a fast and flexible denoising CNN with a tunable noise level map as the input. Motivated by the observation of noise of the estimated noise map (NoN), Ma \textit{et al.}~\cite{ma2021dbdnet} formulate the denoising as a process of reducing NoN from the coarse noise map. To improve the training robustness and model interpretability, Liu \textit{et al.}~\cite{liu2022true} design a few feature extraction streams which are enclosed by wavelet transform and inverse transform. Jia \textit{et al.}~\cite{jia2022pdnet} design to learn a specialized PDE network which depicts the information propagation in the network as a reaction-diffusion–advection process. In addition, based on the framework of Generative adversarial network (GAN), Vo \textit{et al.}~\cite{vo2021hi} utilise three generators by considering the reconstruction of low frequency and high frequency, respectively. However, since the spatially invariant noise distribution deviates from real noise, these models experience significant performance degradation in real image denoising.

\subsection{Real Noise Removal}
To improve the performance in real cases, the research trend progressively focuses on real image denoising. Very recently, some real datasets are publicly available, \textit{e.g.,} SIDD~\cite{sidd}, DND~\cite{dnd} and RN15~\cite{rni}. With the continuous progress, some CNN-based methods implicitly learn the noise prior by extending the methods of synthetic noise removal, which is fully trained on real data. Anwar \textit{et al.}~\cite{ridnet} propose a single-stage blind real image denoising network based on a residual on the residual structure and channel attention. Liu \textit{et al.}~\cite{invdn} transform a noisy image into a low-resolution clean image and a latent representation containing noise which is replaced with another one sampled from a prior distribution during reversion. Since such methods train and test the models in SIDD dataset, they may face the risk of overfitting to noise prior of certain dataset. More blind denoising experiment is conducted which also supports this observation.

Another type of CNN-based methods explicitly estimates the noise prior to perform non-blind image denoising. SOTAs always consider real noise approximated by heteroscedastic Gaussian/Poisson Gaussian distributions with in-camera signal processing pipelines. Such noise assumption is used to generate synthetic noisy images which can provide supervision for explicit noise estimation. By following SOTAs~\cite{guo2019toward,aindnet}, our model is also trained by using both real-world noisy images and synthetic noisy images, which builds a bridge from synthetic noise to real noise. To precisely model the noise distribution, Abdelhamed \textit{et al.}~\cite{abdelhamed2019noise} combine well-established parametric noise models with normalizing flow networks, which is flexible and expressive. Nevertheless, the above SOTAs estimate the noise prior only once which limits the precision of denoising. Accordingly, in the CNN structure, we explicitly introduce alternative noise estimation and non-blind denoising to continuously update the noise prior. In addition, most SOTAs always stack a few convolutional layers to estimate noise. By contrast, based on the property of heteroscedastic Gaussian distribution, the proposed noise estimation adopts iterative affine transform blocks. The scaling and translation of transform blocks can explicitly predict the stationary noise component and the signal-dependent noise component, respectively. To update the estimated noise prior, Yue \textit{et al.}~\cite{vdn} provide another method which integrates both noise estimation and image denoising into an unique Bayesian framework by taking the intrinsic clean image and the noise variances as latent variables. However, the common problem of existing methods is still unresolved. That is, by concatenating the estimated noise features in channel dimension, it cannot adaptively tune the denoising patterns for all feature positions by a convolutional layer with spatially sharing kernels. To the best of our knowledge, based on the analysis on noise prior and frequency details, this work is the first to propose a conditional filter in which convolutional kernels can be adaptively inferred for the local windows centered by all feature positions. Also, sufficient experiments verify the effectiveness of our CFNet for image denoising under variable cases.

\begin{figure*}[t]
    \centering
    \includegraphics[width=0.95\textwidth]{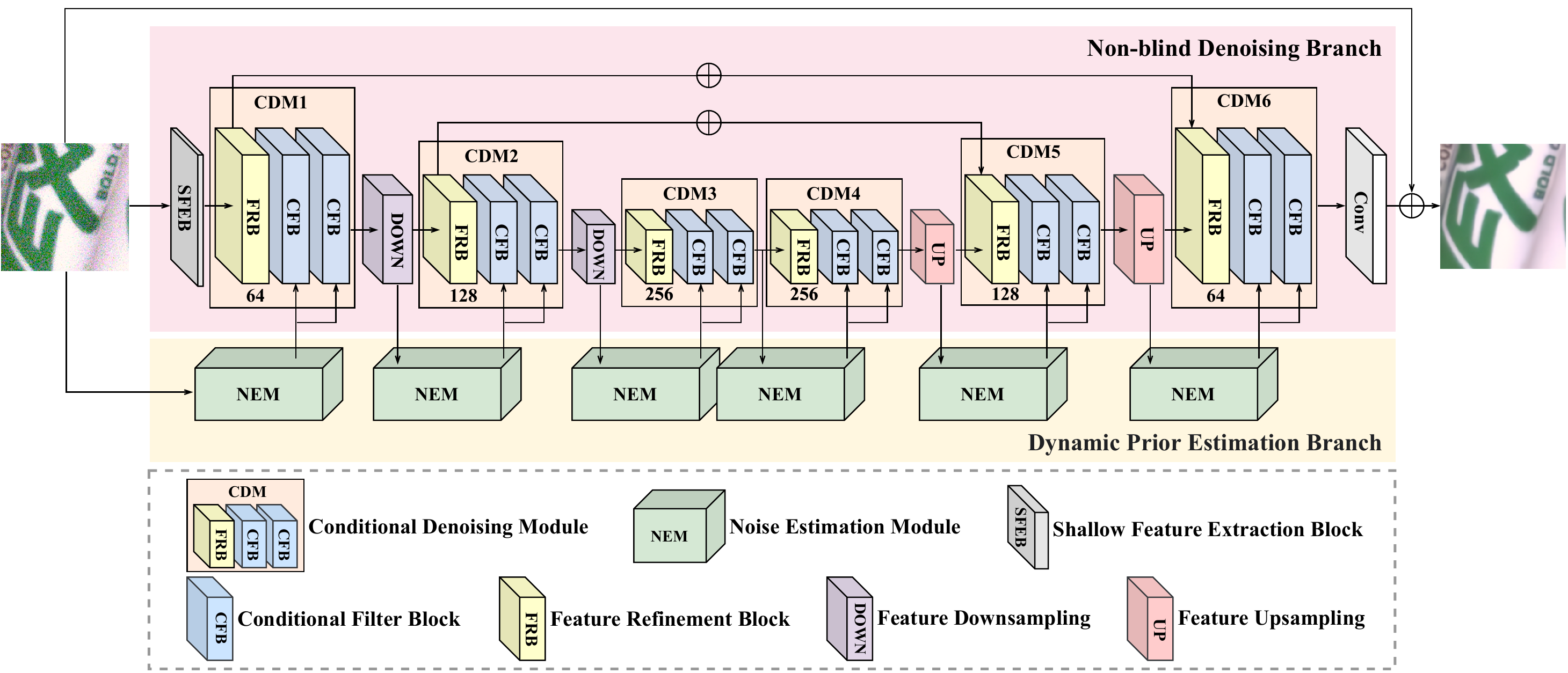}
    \caption{The proposed CFNet consists of two branches, \textit{i.e.,} non-blind denoising and dynamic prior estimation. Particularly, the Noise Estimation Module (NEM) and the Conditional Filter Block (CFB) are introduced in section~\ref{sec:nem} and section~\ref{sec:cfb}, respectively.}
    \label{fig2}
\end{figure*}

\section{The Proposed CFNet}
According to the analysis of features from the image and the noise map, we propose a novel conditional filter to adaptively infer the denoising kernels for local windows, which is further used to construct our CFNet.

\subsection{Overview}
As shown in Eq.~\eqref{eq:overall_eq}, given an image $\mathbf{y}$ corrupted by real noise, the goal of our generative model $CFNet\left(\bm{\theta}|\mathbf{y}\right)$ parameterized by $\bm{\theta}$ is to recover the residual between the noise-free counterpart $\mathbf{x}$ and the input $\mathbf{y}$.
\begin{equation}\label{eq:overall_eq}
\mathbf{x}=CFNet\left(\bm{\theta}|\mathbf{y}\right)+\mathbf{y}   
\end{equation}

The architecture of our CFNet based on U-net~\cite{unet} is shown in Fig.~\ref{fig2}. Unlike static noise estimation~\cite{guo2019toward,aindnet,dncnn}, the proposed CFNet alternatively performs noise estimation and feature denoising in two separate branches, \textit{i.e.,} dynamic prior estimation and non-blind denoising. This design can continuously update the noise prior as iterative feature denoising. The activation function of CFNet is PRelu~\cite{heprelu}.

More specifically, the non-blind denoising branch sequentially includes a shallow feature extraction block (SFEB) consisting of 3 convolutional layers, 6 conditional denoising modules (CDMs) and a convolutional layer. An average pooling layer or a transposed convolutional layer is adopted to link two adjacent conditional denoising modules. Then, by following U-net, symmetrical structure with the channel numbers from 64 to 256 are designed for 6 conditional denoising modules. The feature dimensions are fixed within each conditional denoising module consisting of a feature refinement block (FRB) and $t$ conditional filter blocks (CFBs). The value of $t$ is discussed in the experiment section. A feature refinement block includes five convolutional layers. Moreover, skip connections are introduced between feature refinement blocks in the symmetrical pair of conditional denoising modules which have the same feature dimensions except for the most inner pair.

By following~\cite{guo2019toward,aindnet}, we also use real noisy images and synthetic noisy images corrupted by heteroscedastic Gaussian noise with ISP pipeline to train the proposed CFNet. The synthetic noisy images can provide the supervision for noise estimation. Specifically, the corresponding dynamic prior estimation branch includes 6 noise estimation modules (NEMs) which is the same as the number of conditional denoising modules in the non-blind denoising branch. The details of the noise estimation module and the conditional filter block are explained in section~\ref{sec:nem} and section~\ref{sec:cfb}, respectively.

\subsection{Loss Function}
The loss function has two terms, \textit{i.e.,} $\ell_{asymm}$ and $\ell_{rec}$, which simultaneously supervises noise estimation and non-blind image denoising for synthetic data. In addition, due to the lack of supervision information, only $\ell_{rec}$ is used for real data. The $\ell_{asymm}$ is defined as follows.  
\begin{equation}\label{eq:loss_asymm}
\ell_{asymm} = \sum_{i} \left | \alpha - I_{\bm{\hat{\delta}}(y_{i})-\bm{\delta}(y_{i})}\right |\cdot (\hat{\bm{\delta}}(y_{i})-\bm{\delta}(y_{i}))^2  
\end{equation}
\begin{equation}\label{eq:asymm_e}
I_{e}=\begin{cases}1,\hspace{3mm}&e<0\\0,&otherwise\end{cases} 
\end{equation}
where $\alpha$ is the over-estimation factor in the range from 0 to 0.5~\cite{guo2019toward}, $I_{e}$ is the indicator function, $\bm{\hat{\delta}}(y_{i})$ and $\bm{\delta}(y_{i})$ are the predicted noise prior and the ground truth for arbitrary pixel position $y_{i}$.

As shown in Eq.~\eqref{eq:l1} and Eq.~\eqref{eq:l2}, $\ell_{rec}$ adopts $L_1$ loss and $L_2$ loss for the tasks of real noise removal and synthetic noise removal, respectively. 
\begin{equation}\label{eq:l1}
\ell_{rec}= \left \|CFNet\left(\bm{\theta}|\mathbf{y}\right)+\mathbf{y}-\mathbf{x}  \right \|_1   
\end{equation}
\begin{equation}\label{eq:l2}
\ell_{rec}= \left \|CFNet\left(\bm{\theta}|\mathbf{y}\right)+\mathbf{y}-\mathbf{x}  \right \|_2^2    
\end{equation}
The final loss function is shown in Eq.~\eqref{eq:Loss}.
\begin{equation}\label{eq:Loss}
\ell = \ell_{rec} + \lambda\cdot\ell_{asymm}
\end{equation}
where $\lambda$ balances $\ell_{rec}$ and $\ell_{asymm}$.

\begin{figure*}[t]
    \centering
    \includegraphics[width=0.95\textwidth]{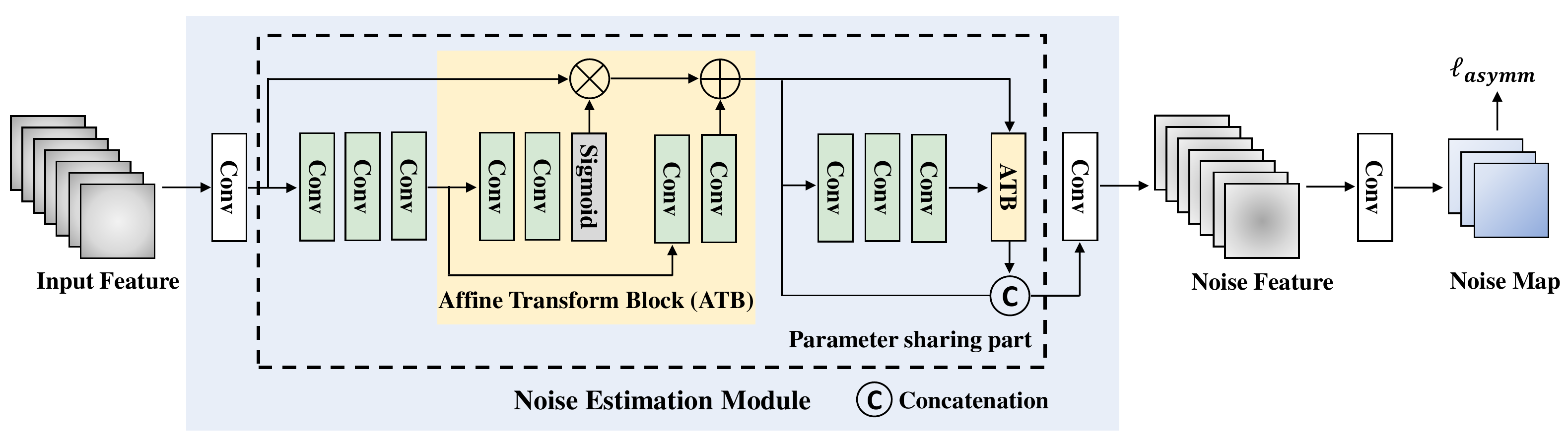}
    \caption{The structure of Noise Estimation Module (NEM). Due to the different channel numbers of input and output among NEMs, the parameters of the first NEM is shared in the following NEMs except two convolutional layers of the module ports. The estimated noise map with three channels is further exploited from the output features by a convolutional layer, which is supervised by $\ell_{asymm}$.}
    \label{fig3}
\end{figure*}

\subsection{Noise Estimation Module}\label{sec:nem}
The details of noise estimation module are shown in Fig.~\ref{fig3}. The input of the $i$-$th$ noise estimation module is the same as the $i$-$th$ conditional denoising module except for the first one whose input is the original noisy image.

As shown in Eq.~\eqref{eq:syn_real_noise}, heteroscedastic Gaussian distribution $\mathbf{n}\left(\mathbf{L}\right)\sim\mathcal N\left(0,\delta^2\left(\mathbf{L}\right)\right)$ followed by ISP pipeline~\cite{guo2019toward} is adopted to generate synthetic signal-dependent noise.
\begin{equation}\label{eq:syn_real_noise}
\bm{\delta^2}\left(\mathbf{L}\right)=\mathbf{L}\cdot\delta_d^2+\delta_s^2 
\end{equation}
where $\mathbf{L}$ is the irradiance image of raw pixels, $\mathbf{L}\cdot\delta_d^2$ and $\delta_s^2$ represent the signal-dependent noise component and the stationary noise component, respectively.

It is observed that the mapping from $\mathbf{L}$ to the noise map $\bm{\delta^2}\left(\mathbf{L}\right)$ is an affine transform parameterized by $\delta_d^2$ and $\delta_s^2$. So, by representing the irradiance image $\mathbf{L}$ in the high-dimension feature domain via stacking convolutional layers, we design affine transform block to learn the affine transform. The corresponding implementation for learning scaling and translation consists of two convolutional layers followed by sigmoid activation and two convolutional layers, respectively. Due to the different channel numbers of input and output among noise estimation modules, a convolutional layer is used for each port of the module. They change the input channel number to 64, and keep the same output dimensions as the features $\mathbf{F_{NDB}^{FR_i}}$ which are the output of the $i$-$th$ feature refinement block in the non-blind denoising branch. Since the supervision is only available in the first noise estimation module, the parameters are shared in the following modules except two convolutional layers of the module ports. That is, by reusing the learned noise estimation function, the noise prior can be continuously exploited in the following stages. Also, we find that two affine transform blocks which are concatenated to obtain the noise features are slightly better than only one affine transform block. The improvement may attribute to ensemble learning in the from-coarse-to-fine framework. The $i$-$th$ estimated noise features $\mathbf{F_{DEB}^i}$ are shared in the two conditional filter blocks of the $i$-$th$ conditional denoising module. The estimated noise map with 3 channels is further extracted from the output features $\mathbf{F_{DEB}^i}$ by a convolutional layer, which is supervised by $\ell_{asymm}$.

\begin{figure*}[t]
    \centering
    \includegraphics[width=0.94\textwidth]{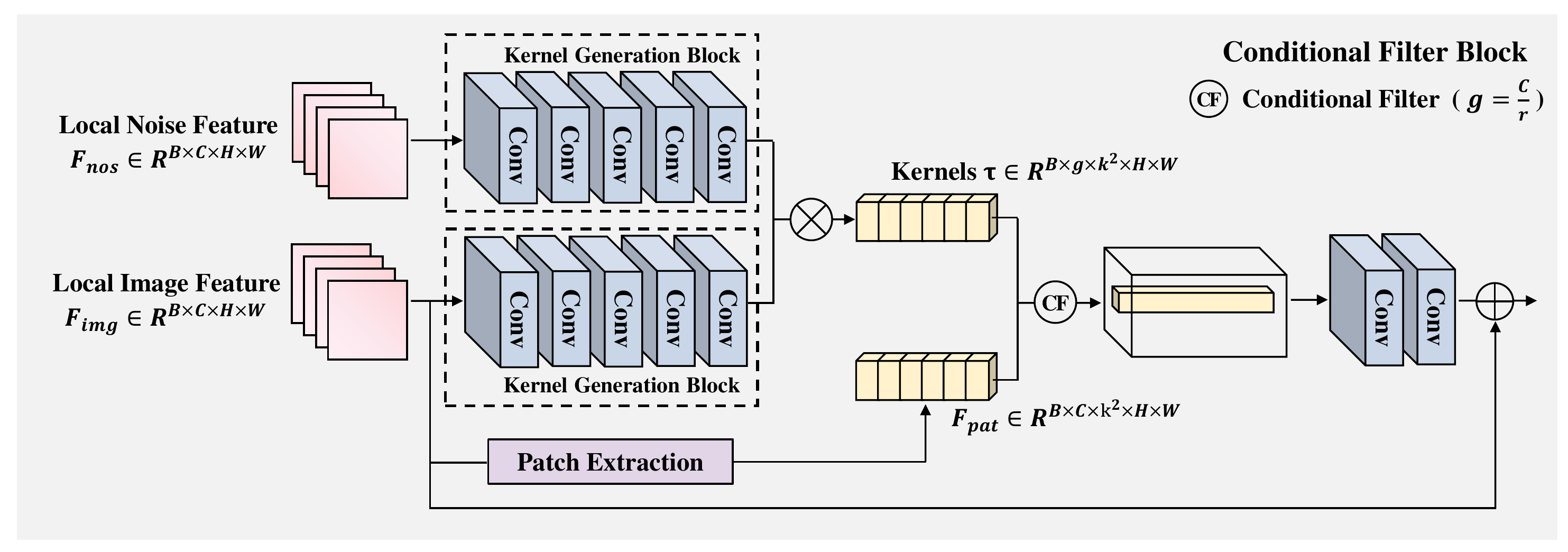}
    \caption{The structure of the Conditional Filter Block. Two light subnetworks with the same structure are designed to model the kernel generation functions for image features and noise features. They are used to generate the conditional denoising kernel via hadmard product. The channel number of conditional filter is $g$ by sharing the kernel for $r$ channels within the local windows of the same spatial positions.}
    \label{fig4}
\end{figure*}

\subsection{Conditional Filter Block}\label{sec:cfb}
As shown in Fig.~\ref{fig4}, a conditional filter block includes a conditional filter and two convolutional layers. Skip connection is used for the image features from the input to the output. The novel conditional filter is defined as Eq.~\eqref{eq:cf}, which infers the denoising kernels within the local windows by feeding the corresponding image features $\mathbf{F_{img}}$ and the noise features $\mathbf{F_{nos}}$. Then, the specifically learned denoising kernels are performed on the corresponding local image features $\mathbf{F_{img}}$ via content-aware convolution.
\begin{equation}\label{eq:cf}
\hat{\mathbf{F}}\mathbf{_{img}}\left(p,c\right)=\sum_{q \in \mathbf{N}\left(p,c\right)}\bm{\mu}\left(q,c\right)\cdot\bm{\gamma}\left(q,c\right)\cdot\mathbf{F_{img}}\left(q,c\right)    
\end{equation}
where $\hat{\mathbf{F}}\mathbf{_{img}}$ and $\mathbf{F_{img}}$ represent the output and the input image features of the conditional filter block, respectively, $p$ and $q$ are the spatial positions, $c$ is the channel index, $\mathbf{N}\left(p, c\right)$ denotes the local window centered by the specific feature position $\left(p, c\right)$, $\bm{\mu}$ and $\bm{\gamma}$ are the corresponding kernels by analysing local features from the image and the noise map, respectively.

We define the proposed conditional filter as a directional feature fusion where the roles of noise features and image features are asymmetric. By contrast, channel-wise feature concatenation is undirectional fusion with symmetric relation which is less interpretable. Obviously, the naive implementation which individually learns all kernel weights in $\bm{\mu}$ and $\bm{\gamma}$ drastically increases the parameters by fixing the input resolution. For flexibility and efficiency, we only learn kernel generation functions instead of individual weights. More specifically, two light subnetworks with the same structure are designed to model the kernel generation functions for the image features and the noise features, respectively. That is, by denoting $B, C, H, W$ as the feature dimensions of batch, channel, height and width, two kernel generation functions learn the mapping from the features $\mathbf{F}\in \mathbf{R^{B\times C\times H\times W}}$ to the corresponding kernels $\bm{\mu}, \bm{\gamma}\in \mathbf{R^{B\times C\times k^2\times H\times W}}$ in the domains of image features and noise features. So, for each feature position $\left(c,h,w\right)$ with $c\in C, h\in H, w\in W$, $k^2$ weights in the conditional kernel $\bm{\tau}=\bm{\mu}\otimes\bm{\gamma}$ are generated by hadmard product. As illustrated in Fig.~\ref{fig4}, we only use five convolutional layers to model each kernel generation function.

Motivated by Involution~\cite{involution}, we also share the kernel of conditional filter for $r$ channels within the local windows of the same spatial position to compact the model. Therefore, the dimensions of each kernel is $\bm{\tau}\in \mathbf{R^{B\times g\times k^2\times H\times W}}$ with the channel group number $g=\frac{C}{r}$. According to the channel numbers are different among the conditional denoising modules based on U-net, we fix the channel group number $g$ instead of the channel number of kernel sharing $r$ in all conditional filter blocks.

\section{Experiments}
In this section, we first introduce the implementation details and the datasets for testing. Moreover, the qualitative and quantitative evaluation of our CFNet is provided compared with SOTAs.

\subsection{Implementation Details}
To train our model for synthetic noise removal, the training data is generated by adding Additive White Gaussian Noise (AWGN) to the clean images from BSD~\cite{bsddatabase} and Div2K~\cite{div2k} datasets. We consider two types of model training. One is non-blind denoising which trains the specific models for certain standard deviations of noise, \textit{i.e.,} 15, 25 and 50. The other is blind denoising which trains a universal model for a standard deviation range of $[0, 55]$. For real noise removal, the proposed CFNet is jointly trained on 3859 synthetic noisy images from Waterloo dataset~\cite{waterloo} and 320 real noisy images from SIDD dataset~\cite{sidd}. More specifically, the synthetic noisy images are generated by the noise model adopted in~\cite{guo2019toward}. All the training data consists of the patches extracted from the corresponding images which are further augmented by random rotation, cropping, and flipping. The batch sizes are 32 and 10 for synthetic noise removal and real noise removal, respectively.

In addition, the number of conditional filter blocks $t$ is 2. The overestimation factor $\alpha$ is set to 0.35 and 0.25 for synthetic noise removal and real noise removal, respectively. $3\times3$ kernels are adopted for all standard convolutional layers except affine transform blocks in which kernel size is $1\times1$. The kernel size $k$ and the channel number $g$ of conditional filters are 3 and 16, respectively. Furthermore, since the precision of noise estimation is improved as the model training, the balance coefficient $\lambda$ is progressively decreased to focus on image denoising. Adam optimizer~\cite{Adam} is used with the default parameters whose learning rate is decayed by cosine annealing. Specifically, we set the initial values of the learning rate and $\lambda$ to $5\times10^{-5}$ and 0.5. They are simultaneously halved by every $1\times10^{6}$ and $1\times10^{5}$ iterations for synthetic noise removal and real noise removal, respectively. Our CFNet is implemented by the Pytorch package and trained on an NVIDIA 3090 GPU. For testing, Peak Signal-to-Noise Ratio (PSNR) and Structural Similarity Index Measure (SSIM) are used as the evaluation metrics.

\subsection{Test Datasets}
We test our CFNet for removing synthetic AWGN noise with specific levels of 15, 25 and 50 on two grayscale datasets \textit{i.e.,} Set12~\cite{dncnn} and BSD68~\cite{bsd68}, and three color datasets \textit{i.e.,}  Kodak24~\cite{kodak}, CBSD68~\cite{bsd68} and McMaster~\cite{mcmater}. It should be noted that the scenes of BSD68 dataset and CBSD68 dataset are the same. In addition, three real datasets, \textit{i.e.,} SSID~\cite{sidd}, DND~\cite{dnd} and RNI15~\cite{rni}, are used to evaluate the performance of real noise removal. More description of real datasets which are recently public are expressed as below.

\begin{itemize}
\item SSID~\cite{sidd} is obtained by five smartphone cameras with small apertures and sensor sizes, which has 30000 pairs of real noisy images and clean ground truth. We use 1280 images cropped from 40 images for validation. Since the noise-free images are not fully public, the results are evaluated via the online system.
\item DND~\cite{dnd} is captured by four consumer-grade cameras
of different sensor sizes, which includes 50 real noisy images with the corresponding noise-free counterparts. Then, 1000 sub-images with resolution of 512$\times$512 are cropped. Also, the results are reported through the online submission system.
\item RNI15~\cite{rni} provides 15 real noisy images without ground truth. So, only visual comparisons is provided.
\end{itemize}

\begin{table}[!t]\footnotesize
	\centering
	\caption{Average PSNR of the test images from Set12 and BSD68 grayscale datasets with AWGN noise levels of 15, 25 and 50. Top Subtable: The non-blind denoising models are specifically trained for certain noise levels. Bottom Subtable:The blind denoising models are universally trained for a range of noise levels. ({\color{red}red} is the optimal result, {\color{blue} blue} is the sub-optimal result)}  
	\label{tab:synthetic_noise}
	\begin{tabular}{cccccccccc}
	    \hline
		\multirow{2}*{Method} & \multicolumn{2}{c}{15} &
		\multicolumn{2}{c}{25} & \multicolumn{2}{c}{50} \\ 
		\cline{2-7}
		&Set12&BSD68&Set12&BSD68&Set12&BSD68\\
	    \hline 
		BM3D~\cite{bm3d} &32.37&31.07&29.97&28.57&26.72&25.62\\
		DnCNN~\cite{dncnn}&32.86&31.73&30.44&29.23&27.18&26.23\\
		FFDNet~\cite{ffdnet}&32.75&31.63&30.43&29.19&27.32&26.29 \\
		RIDNet~\cite{ridnet}&32.91&31.81&30.60&29.34&27.43&26.40\\
		AINDNet~\cite{aindnet}&32.92&31.69&30.61&29.26& 27.51& 26.32\\
		ADNet~\cite{adnet}&32.98&31.74&30.58&29.25&27.37&26.29\\
		BRDNet~\cite{brdnet}&33.03& 31.79& 30.61&29.29&27.45&26.36 \\
		DudeNet~\cite{dudenet}&32.94&31.78&30.52&29.29&27.30&26.31 \\
		DNA-Net~\cite{dnanet}&{\color{red}33.14}&{\color{red}31.87}&{\color{blue}30.83}&{\color{blue}29.41}&27.77&26.52 \\
		DBDNet~\cite{ma2021dbdnet}&33.03& 31.85& 30.65&29.37&27.46&26.43 \\
		DIPD~\cite{dipdiccv21}&32.26&31.21&29.76&28.78&26.47&25.81 \\
		APDNet~\cite{apdnets}&32.96&31.79&30.74&29.35&27.78&{\color{red}26.58} \\
		DeamNet~\cite{deamnet}&33.10&31.83&30.82&29.39&{\color{blue}27.85}&26.51 \\
		\hline
	    CFNet  & {\color{blue}33.13} &{\color{blue}31.85} &{\color{red}30.86}  &{\color{red}29.44}&{\color{red}27.88}&{\color{blue}26.57} \\
	    \hline
	    DnCNN-B~\cite{dncnn}&32.68&31.61&30.36&29.16&27.21&26.23\\
		FFDNet-B~\cite{ffdnet}&32.75&31.63&30.43&29.19&27.32&26.29 \\
	    IRCNN-B~\cite{ircnn}&32.76&31.63&30.37&29.15&27.12&26.19 \\
		ADNet-B~\cite{adnet}&32.77&31.56&30.46&29.14&27.33&26.23\\
		DBDNet-B~\cite{ma2021dbdnet}&32.90&  31.77& 30.53&29.25&27.44& 26.38 \\
		DeamNet-B~\cite{deamnet}&{\color{blue}33.03}&{\color{blue}31.80}&{\color{blue}30.75}&{\color{blue}29.36}&{\color{blue}27.63}&{\color{blue}26.43} \\
	    \hline
	    {CFNet-B}  & {\color{red}33.10} &{\color{red}31.83} &{\color{red}30.81}  &{\color{red}29.43}&{\color{red}27.74}&{\color{red}26.51} \\
	    \hline
	\end{tabular}
\end{table}

\begin{figure}[t]
\setlength{\abovecaptionskip}{0pt} 
	\centering
	\subfigure[Noisy]{\centering
	\begin{minipage}[b]{0.16\textwidth}
    \includegraphics[width=1\linewidth]{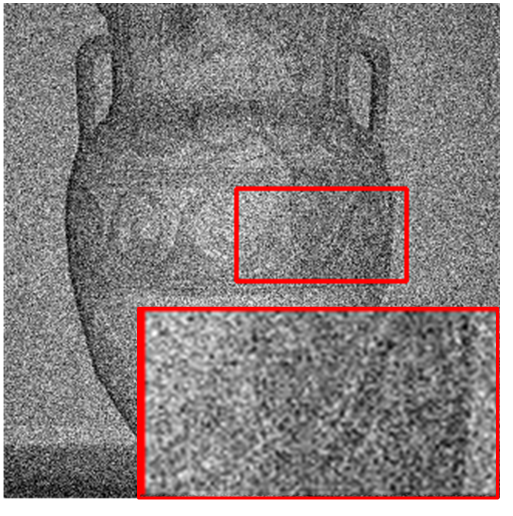}
    \includegraphics[width=1\linewidth]{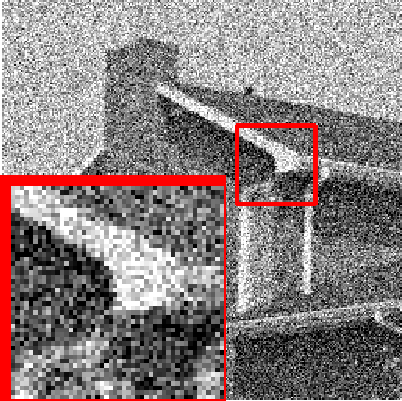}
    \end{minipage}}
	\subfigure[DnCNN]{\centering
	\begin{minipage}[b]{0.16\textwidth}
    \includegraphics[width=1\linewidth]{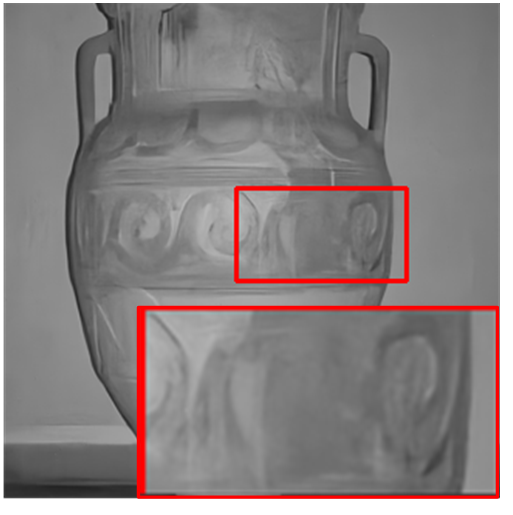}
    \includegraphics[width=1\linewidth]{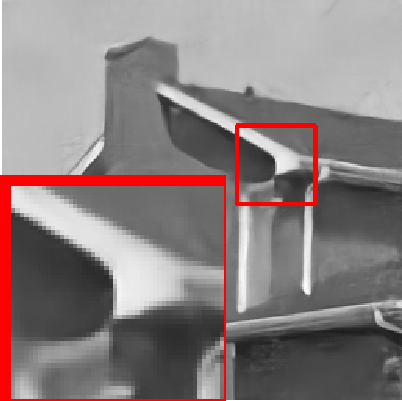}
    \end{minipage}}
	\subfigure[RIDNet]{\centering
	\begin{minipage}[b]{0.16\textwidth}
    \includegraphics[width=1\linewidth]{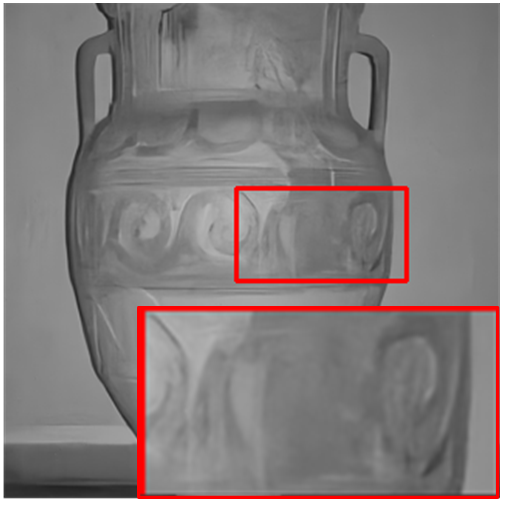}
    \includegraphics[width=1\linewidth]{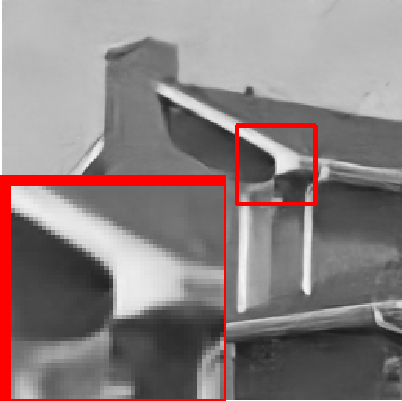}
    \end{minipage}}
	\subfigure[ADNet]{\centering
	\begin{minipage}[b]{0.16\textwidth}
    \includegraphics[width=1\linewidth]{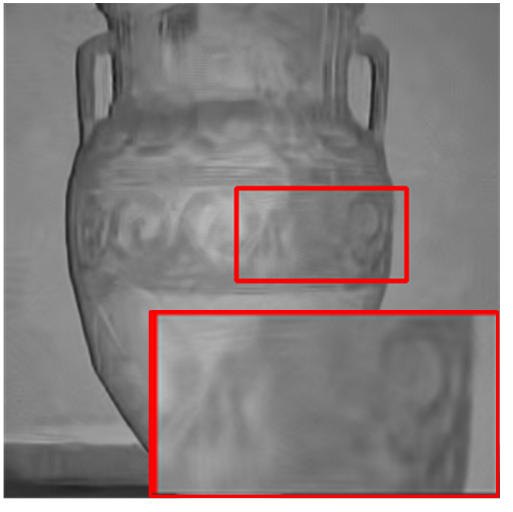}
    \includegraphics[width=1\linewidth]{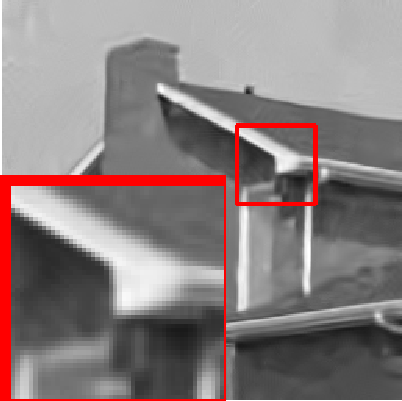}
    \end{minipage}}	
	\subfigure[\bf{Ours}]{\centering
	\begin{minipage}[b]{0.16\textwidth}
    \includegraphics[width=1\linewidth]{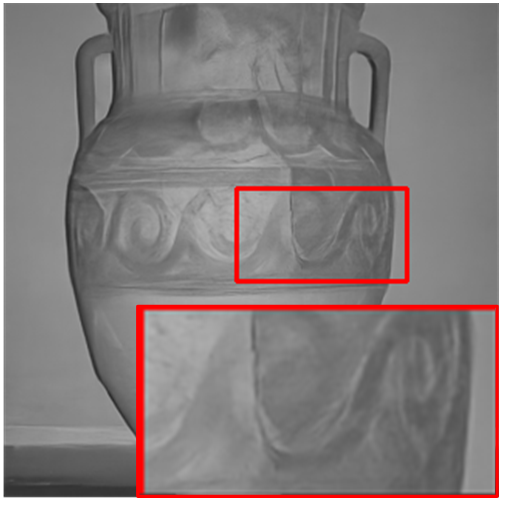}
    \includegraphics[width=1\linewidth]{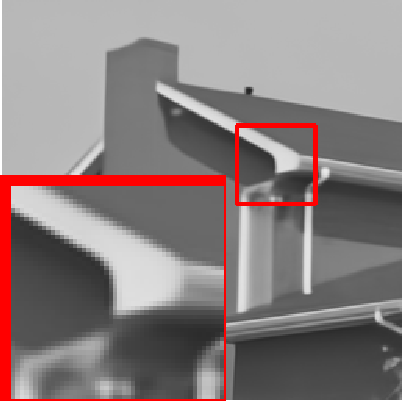}
    \end{minipage}}	
	\subfigure[GT]{\centering
	\begin{minipage}[b]{0.16\textwidth}
    \includegraphics[width=1\linewidth]{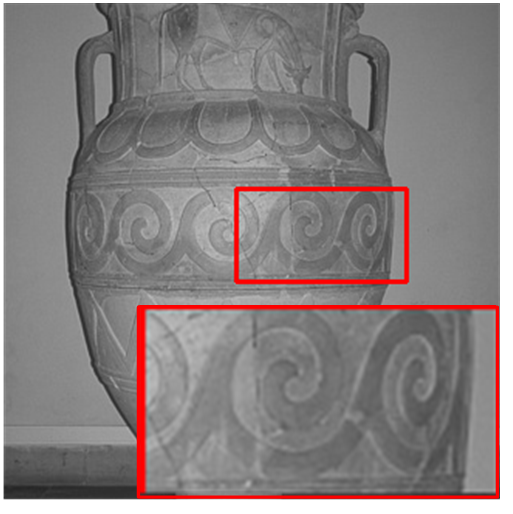}
    \includegraphics[width=1\linewidth]{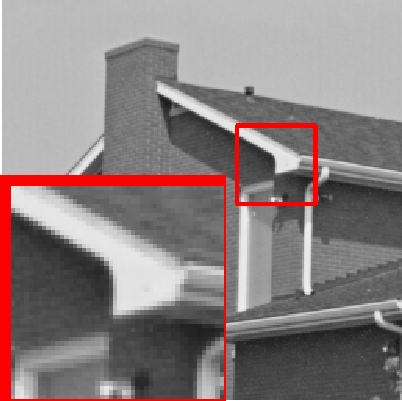}
    \end{minipage}}	
	\caption{Visual comparison of the results from synthetic grayscale datasets with AWGN noise level of 50. The first and second rows are the results from Set68 and Set12 datasets, respectively. According to the (f) ground truth, the noisy image is denoised by (b) DnCNN, (c) RIDNet , (d) ADNet and  (e) Our CFNet.}
	\label{fig_Synthetic}
\end{figure}

\subsection{Result for Synthetic Noise Removal}
The performance of the proposed CFNet is first evaluated on two synthetic grayscale datasets, \textit{i.e.,} Set12 and BSD68, and three synthetic color datasets, \textit{i.e.,} CBSD68, Kodak24 and McMaster. Generally, two types of model training for non-blind denoising and blind denoising are conducted. They are specifically trained for certain noise levels, \textit{i.e.,} 15, 25 and 50, and universally trained for a range of noise levels [0, 55], respectively. All the results are referred from the original papers except for DeamNet~\cite{deamnet}. For the comparison on all synthetic datasets and denoising modes of this part which are not fully considered in DeamNet, we reproduce the results of DeamNet by the code from the authors. Specifically, DeamNet is trained using the same training dataset and patch size as ours.

\subsubsection{Synthetic Grayscale Datasets}
First, we investigate the performance of non-blind denosing models trained for specific noise levels. 13 SOTAs and benchmarks are compared which includes 2 traditional methods, \textit{i.e.,} BM3D~\cite{bm3d},  DIPD~\cite{dipdiccv21} and 11 CNN-based methods, \textit{i.e.,} DnCNN~\cite{dncnn}, FFDNet~\cite{ffdnet}, RIDNet~\cite{ridnet}, AINDNet~\cite{aindnet}, ADNet~\cite{adnet}, BRDNet~\cite{brdnet}, DudeNet~\cite{dudenet}, DNA-Net~\cite{danet}, DBDNet~\cite{ma2021dbdnet}, APDNet~\cite{apdnets} and DeamNet~\cite{deamnet}. The average PSNR of the competing methods is shown in Tab.~\ref{tab:synthetic_noise}. It is observed that our CFNet reaches the highest average PNSR in the most cases. When the noise level is 15, our model shows comparable result to DNA-Net. Also, the visual comparison is illustrated in Fig.~\ref{fig_Synthetic} which are two examples from BSD68 and Set12 datasets under the noise level of 50. According to the highlighted details, it is shown that DnCNN, RIDNet and ADNet cannot maintain tiny image structure, \textit{e.g.,} texture of the vase. By contrast, thanks to conditional filters which adaptively tune the denoising strength, our CFNet can preserve edges and details as effective denoising.

Then, to further validate the generalization of the proposed CFNet, the blind denoising model CFNet-B is considered to handle a range of noise levels, \textit{i.e.,} from 0 to 55. That is, we add AWGN with standard deviations in the range of $[0, 55]$ to training data and tested the performance for 3 certain noise levels, \textit{i.e.,} 15, 25 and 50. Our blind denoising model CFNet-B is compared with 6 CNN-based methods, \textit{i.e.,} DnCNN~\cite{dncnn}, FFDNet~\cite{ffdnet}, IRCNN~\cite{ircnn}, ADNet~\cite{adnet}, DBDNet~\cite{ma2021dbdnet} and DeamNet~\cite{deamnet}. The average PSNR of the competing methods is shown in Tab.~\ref{tab:synthetic_noise}. Thanks to dynamic noise estimation and adaptive feature denoising, the proposed CFNet-B can reach comparable result to non-blind counterpart CFNet. However, the methods which implicitly learn the noise prior in real dataset, \textit{ADNet-B} perform quite lower than the corresponding non-blind counterparts, respectively. So, they may face the risk of overfitting to noise prior of certain dataset. Compared with SOTAs, the proposed CFNet-B shows improvement by significant margins.

\begin{table}[t]\footnotesize
	\centering
	\caption{Average PSNR of test images from CBSD68, Kodak24 and McMaster color datasets with AWGN noise levels of 15, 25, and 50. Non-blind denoising models and blind denoising models are trained for certain noise levels and a range of noise level, respectively. ({\color{red}red} is the optimal result, {\color{blue} blue} is the sub-optimal result)}  
	\label{tab:color_awgn_denoising}
    \resizebox{\linewidth}{!}{
    \setlength\tabcolsep{0.5pt}
	\begin{tabular}{cccccccccc}
	    \hline
		\multirow{2}*{Method} & \multicolumn{3}{c}{15} &
		\multicolumn{3}{c}{25} & \multicolumn{3}{c}{50} \\ 
		\cline{2-10}
		& CBSD68 & Kodak24 & McMaster & CBSD68 & Kodak24 & McMaster & CBSD68 & Kodak24 & McMaster \\
	    \hline 
		BM3D~\cite{bm3d} & 33.52 & 34.28 & 34.06 & 30.71 & 31.68 & 31.66 & 27.38 & 28.46 & 28.51 \\
		DnCNN~\cite{dncnn}  & 33.98 & 34.73 & 34.80 & 31.31 & 32.23 & 32.47 & 28.01 & 29.02 & 29.21 \\
		FFDNet~\cite{ffdnet}  & 33.80 & 34.55 & 34.47 & 31.18 & 32.11 & 32.25 & 27.96 & 28.99 & 29.14 \\
		IRCNN ~\cite{ircnn}  & 33.86 & 34.56 & 34.58 & 31.16 & 32.03 & 32.18 & 27.86 & 28.81 & 28.91 \\
		ADNet~\cite{adnet}  & 33.99 & 34.76 & 34.93 & 31.31 & 32.26& 32.56 & 28.04 & 29.10 & 29.36 \\
		APDNet~\cite{apdnets}& 33.19 & -- & -- & 31.06 & --& -- & 28.30 & -- & -- \\
		DeamNet~\cite{deamnet}  & {\color{red}34.26} & 34.88 & 35.01 & 31.57 & 32.55 & {\color{blue}32.85} & 28.36 & 29.51 & {\color{blue}29.84} \\
		\hline
	    CFNet  & 34.22 & {\color{blue}34.89}& {\color{red}35.07} & {\color{red}31.64} & {\color{red}32.56} & {\color{red}32.91} & {\color{red}28.45} & {\color{red}29.59} & {\color{red}29.86} \\
	    CFNet-B & {\color{blue}34.25} & {\color{red}34.94}& {\color{red}35.07} & {\color{blue}31.62} & {\color{red}32.56} & 32.84 & {\color{blue}28.42} &{\color{blue}29.54} & 29.82\\
	    \hline
	\end{tabular}}
\end{table}

\begin{figure*}[t]
\setlength{\abovecaptionskip}{0pt} 
	\centering
	\subfigure[Noisy]{\centering
	\begin{minipage}[b]{0.16\textwidth}
    \includegraphics[width=1\linewidth]{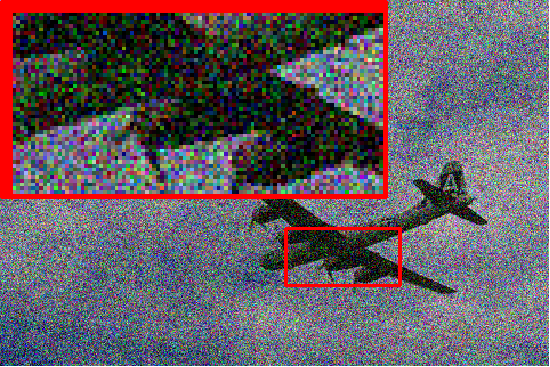}
    \includegraphics[width=1\linewidth]{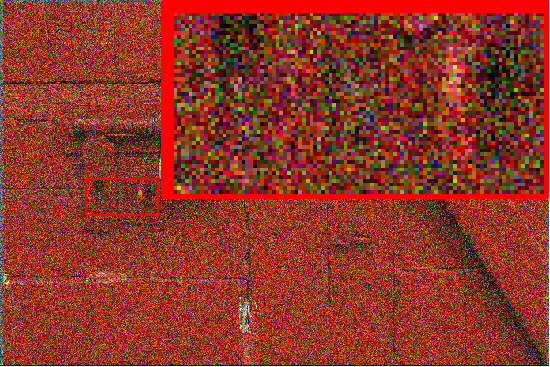}
    \includegraphics[width=1\linewidth]{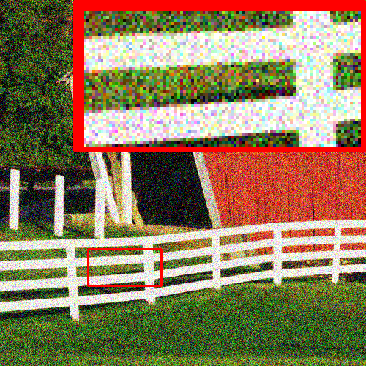}
    \end{minipage}}
	\subfigure[BM3D]{\centering
	\begin{minipage}[b]{0.16\textwidth}
    \includegraphics[width=1\linewidth]{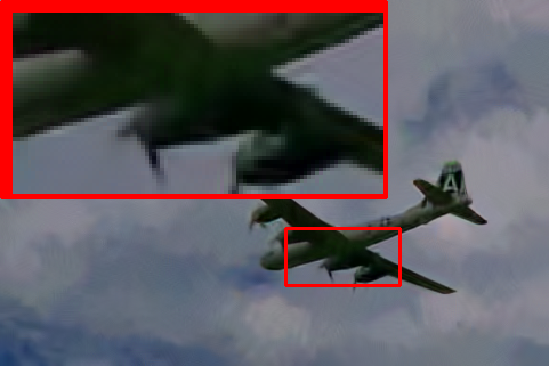}
    \includegraphics[width=1\linewidth]{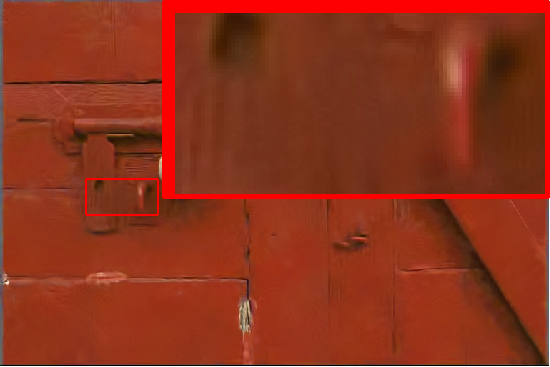}
    \includegraphics[width=1\linewidth]{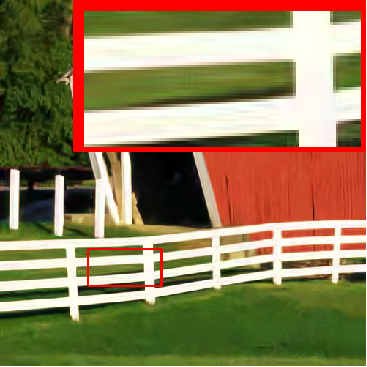}
    \end{minipage}}
	\subfigure[DnCNN]{\centering
	\begin{minipage}[b]{0.16\textwidth}
    \includegraphics[width=1\linewidth]{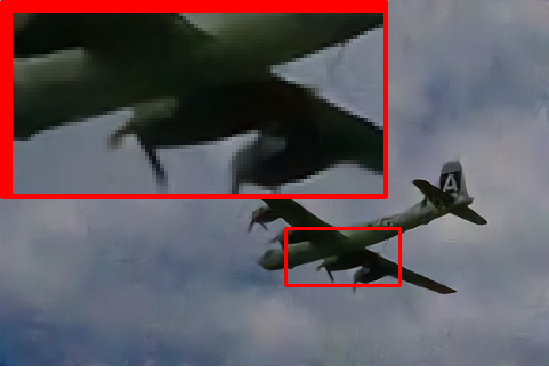}
    \includegraphics[width=1\linewidth]{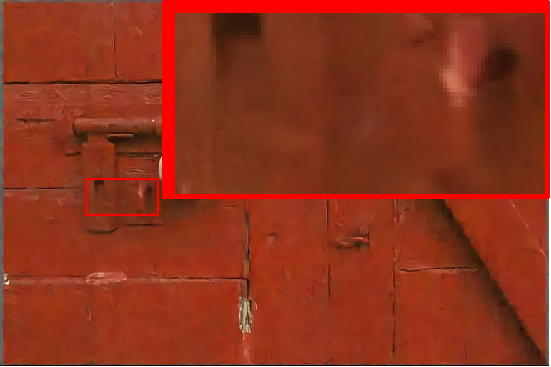}
    \includegraphics[width=1\linewidth]{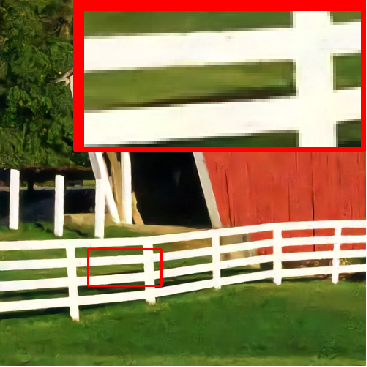}
    \end{minipage}}
	\subfigure[DeaMet]{\centering
	\begin{minipage}[b]{0.16\textwidth}
    \includegraphics[width=1\linewidth]{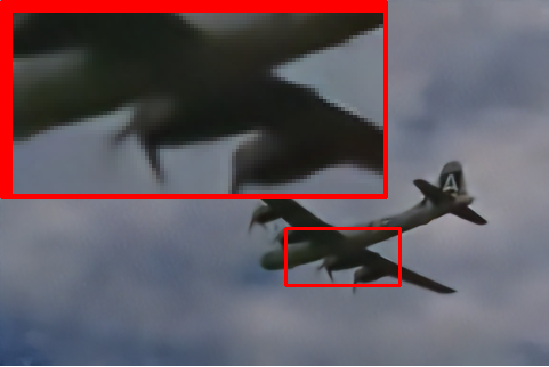}
    \includegraphics[width=1\linewidth]{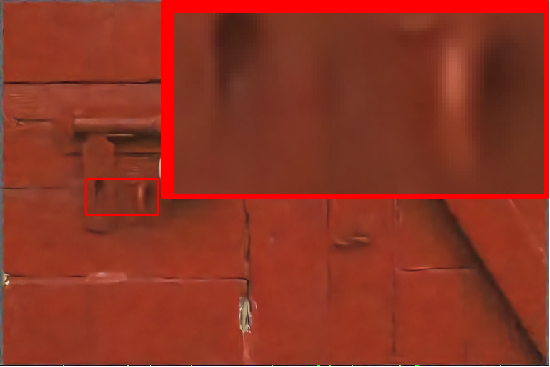}
    \includegraphics[width=1\linewidth]{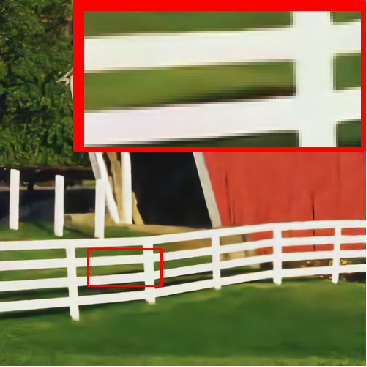}
    \end{minipage}}	
	\subfigure[\bf{Ours}]{\centering
	\begin{minipage}[b]{0.16\textwidth}
    \includegraphics[width=1\linewidth]{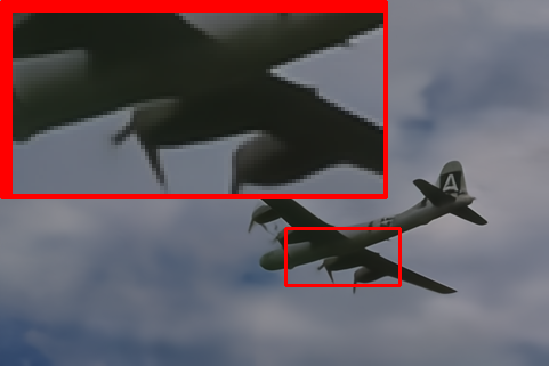}
    \includegraphics[width=1\linewidth]{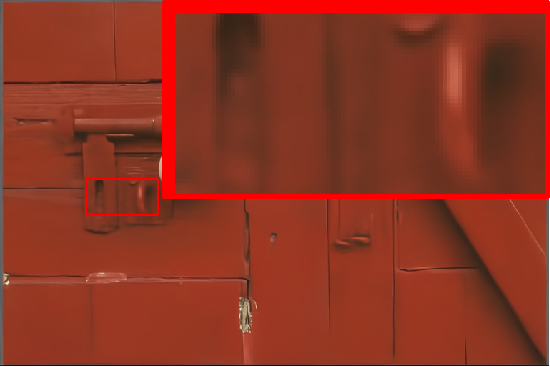}
    \includegraphics[width=1\linewidth]{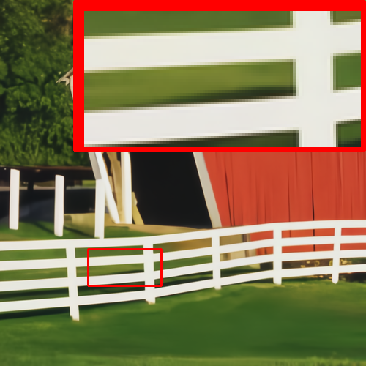}
    \end{minipage}}	
	\subfigure[GT]{\centering
	\begin{minipage}[b]{0.16\textwidth}
    \includegraphics[width=1\linewidth]{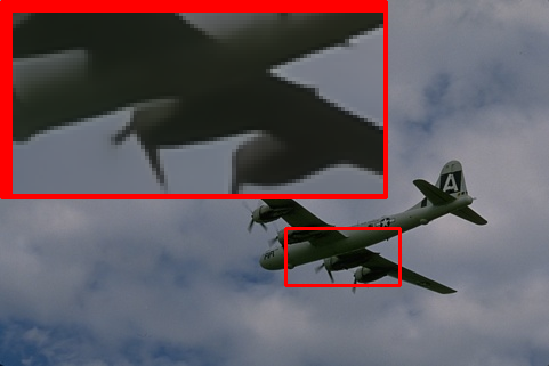}
    \includegraphics[width=1\linewidth]{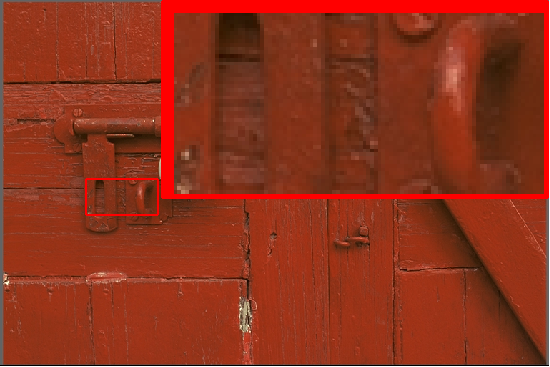}
    \includegraphics[width=1\linewidth]{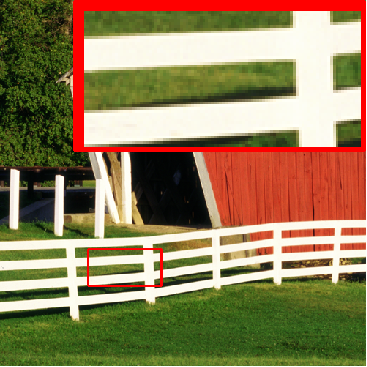}
    \end{minipage}}	
	\caption{Visual comparison of the results from synthetic color datasets with AWGN noise level of 50. The first, second and third rows are the results from CBSD68, Kodak24 and McMaster datasets, respectively. According to the (f) ground truth, the (a) noisy image is denoised by (b) BM3D, (c) DnCNN, (d) DeamNet and (e) Our CFNet.}
	\label{fig_AWGN_color_CBSD68}
\end{figure*}

\subsubsection{Synthetic Color Datasets}
We also test the performance of CFNet and CFNet-B on synthetic color datasets, \textit{i.e.,} CBSD68, Kodak24 and McMaster. our CFNet or CFNet-B are compared with the traditional method BM3D~\cite{bm3d} and 6 CNN-based methods, \textit{i.e.,} DnCNN~\cite{dncnn}, FFDNet~\cite{ffdnet}, IRCNN~\cite{ircnn}, ADNet~\cite{adnet}, APDNet~\cite{apdnets} and DeamNet~\cite{deamnet} which are all trained in non-blind mode. The corresponding results are listed in Tab.~\ref{tab:color_awgn_denoising}. Overall, our models can obtain the highest PSNR for the most cases. Especially, the proposed CFNet-B outperforms most SOTAs which are trained on non-blind mode. Furthermore, the performance of the proposed CFNet and CFNet-B is comparable. Interestingly, since the less challenge is provided when the noise level is only 15, the noise estimation has enough precision. In this case, CFNet-B even occasionally outperform CFNet, which reveals that our CFNet is not overfitting to certain noise levels. Instead, the proposed CFNet-B is not confused with various noise levels by efficiently noise estimation and conditional filtering. So, the proposed CFNet-B can provide an efficient way for blind image denoising. The qualitative results are visualized in Fig.~\ref{fig_AWGN_color_CBSD68} for CBSD68~\cite{bsd68}, Kodak24~\cite{kodak} and McMaster~\cite{mcmater} with the noise level of 50, respectively. Compard with SOTAs, our CFNet can recover more details which are closest to the ground truth.

\begin{table}[!t]\footnotesize
\setlength{\abovecaptionskip}{0pt}
\centering %
\caption{Average PSNR of test images from SIDD dataset. ({\color{red} red:} the optimal result, {\color{blue} blue:} the sub-optimal result)\\[3pt]} %
\label{table_sidd} %
\setlength{\tabcolsep}{1mm}{
\begin{tabular}{ccccc}
\hline
Method&Blind& Training&PSNR&SSIM\\
\hline
BM3D~\cite{bm3d}&&-&25.65&0.685\\
KSVD~\cite{ksvd}&&-&26.88&0.842\\
WNNM~\cite{wnnm}&&-&25.78&0.809\\
DnCNN~\cite{dncnn}&\checkmark& Synthetic&23.66&0.583\\
RIDNet~\cite{ridnet}&\checkmark&Real&38.71&--\\
VDN~\cite{vdn}&\checkmark&Real&39.26&0.955\\
GradNet~\cite{gradnet}&\checkmark&Real&38.34&0.946\\
DANet~\cite{danet}&\checkmark&Real&39.25&0.955\\
InvDN~\cite{invdn}&\checkmark&Real&39.28&0.955\\
HI-GAN~\cite{higan}&\checkmark&Real&38.88&0.952\\
DualBDNet~\cite{du2020blind}&\checkmark&Real&38.01&0.943\\
CPNet~\cite{li2021cross}&\checkmark&Real&38.34&{\color{red} 0.957}\\
GreneNet~\cite{grenenet}&\checkmark&Real&{\color{red}39.42}&{\color{red}0.957}\\
DeamNet~\cite{deamnet}&\checkmark&Real&{\color{blue} 39.35}&0.955\\
\hline
CBDNet (All)~\cite{guo2019toward}&\checkmark&All&33.28&0.868\\
CBDNet (All-R)~\cite{guo2019toward}&\checkmark&All&33.47&0.870\\
AINDNet (All)~\cite{aindnet}&\checkmark&All&38.95&0.950\\
AINDNet (All-R)~\cite{aindnet}&\checkmark&All&39.04&0.953\\
\hline
CFNet (All)&\checkmark&All&39.09&0.953\\
CFNet (All-R)&\checkmark&All& 39.34&0.955\\
\hline
\end{tabular}}
\end{table}

\begin{table}[!t]\footnotesize
\setlength{\abovecaptionskip}{0pt}
\centering 
\caption{Average PSNR of test images from DND dataset. ({\color{red} red:} the optimal result, {\color{blue} blue:} the sub-optimal result)\\[3pt]} %
\label{table_dnd} %
\setlength{\tabcolsep}{1mm}{
\begin{tabular}{ccccc}
\hline
Method&Blind& Training&PSNR&SSIM\\
\hline
BM3D~\cite{bm3d}&&-& 34.51 &0.8507\\
KSVD~\cite{ksvd}&&-& 36.49&0.8978\\
WNNM~\cite{wnnm}&&-& 34.67&0.8646\\
DnCNN~\cite{dncnn}&\checkmark& Synthetic& 32.43&0.7900\\
RIDNet~\cite{ridnet}&\checkmark&Real&39.26&0.9528\\
VDN~\cite{vdn}&\checkmark&Real&39.38&0.9518\\
GradNet~\cite{gradnet}&\checkmark&Real&39.44&0.9543\\
DANet~\cite{danet}&\checkmark&Real&39.58&0.9545\\
InvDN~\cite{invdn}&\checkmark&Real&39.57&0.9522\\
HI-GAN~\cite{higan}&\checkmark&Real&39.37&0.9542\\
GreneNet~\cite{grenenet}&\checkmark&Real&39.76&0.9559\\
DAGL~\cite{dagl}&\checkmark&Real&39.83&{\color{blue}0.9570}\\
DBDNet~\cite{ma2021dbdnet}&\checkmark&Real&39.61&--\\
MPRNet~\cite{mprnet}&\checkmark&Real&39.82&0.9540\\
MIRNet~\cite{zamirmirnet}&\checkmark&Real&{\color{blue}39.88}&0.9563\\
DeamNet~\cite{deamnet}&\checkmark&Real&39.63&0.9531\\
\hline
CBDNet (All)~\cite{guo2019toward}&\checkmark&All&38.06&0.9421\\
CBDNet (All-R)~\cite{guo2019toward}&\checkmark&All&38.00&0.9420\\
AINDNet (All)~\cite{aindnet}&\checkmark&All& 39.37& 0.9520\\
AINDNet (All-R)~\cite{aindnet}&\checkmark&All&39.21&0.9518\\
\hline
CFNet (All)&\checkmark&All&{\color{red} 39.92}&{\color{red} 0.9593}\\
CFNet (All-R)&\checkmark&All&39.65&0.9529\\
\hline
\end{tabular}}
\end{table}

\begin{figure*}[!t]
\setlength{\abovecaptionskip}{0pt} 
	\centering
		\centering
	\subfigure[Noisy]{\centering
	\begin{minipage}[b]{0.16\textwidth}
    \includegraphics[width=1\linewidth]{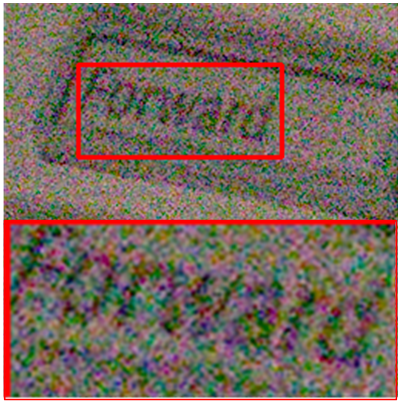}
    \includegraphics[width=1\linewidth]{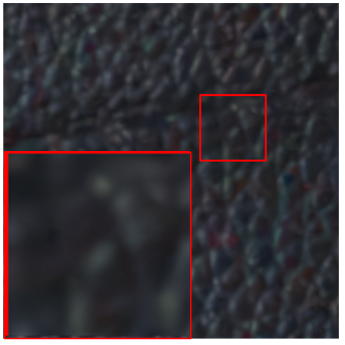}
    \includegraphics[width=1\linewidth]{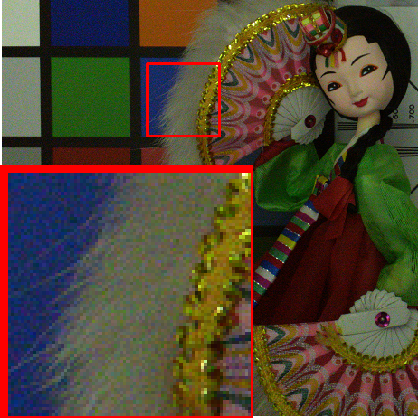}
    \end{minipage}}
	\subfigure[CBDNet]{\centering
	\begin{minipage}[b]{0.16\textwidth}
    \includegraphics[width=1\linewidth]{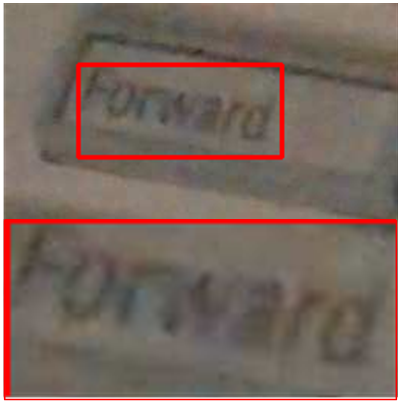}
    \includegraphics[width=1\linewidth]{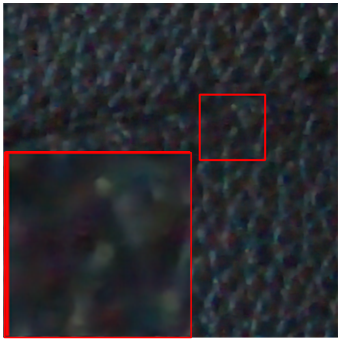}
    \includegraphics[width=1\linewidth]{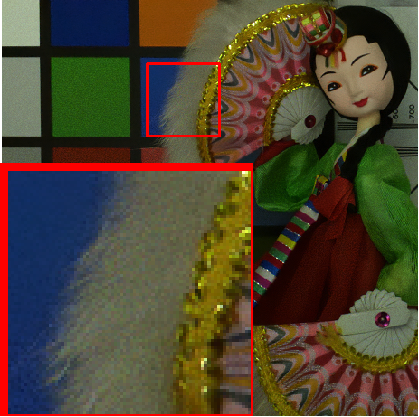}
    \end{minipage}}
	\subfigure[RIDNet]{\centering
	\begin{minipage}[b]{0.16\textwidth}
    \includegraphics[width=1\linewidth]{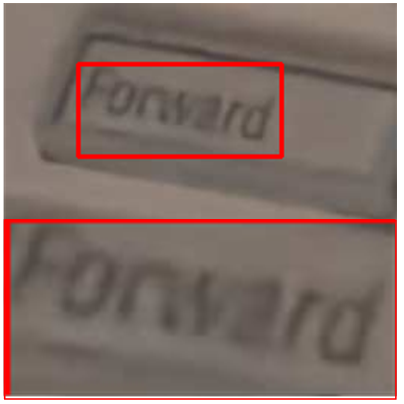}
    \includegraphics[width=1\linewidth]{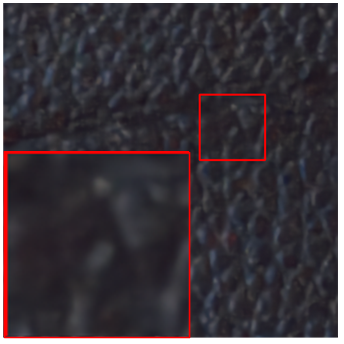}
    \includegraphics[width=1\linewidth]{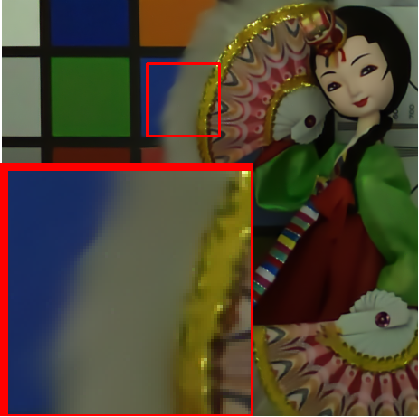}
    \end{minipage}}
	\subfigure[VDN]{\centering
	\begin{minipage}[b]{0.16\textwidth}
    \includegraphics[width=1\linewidth]{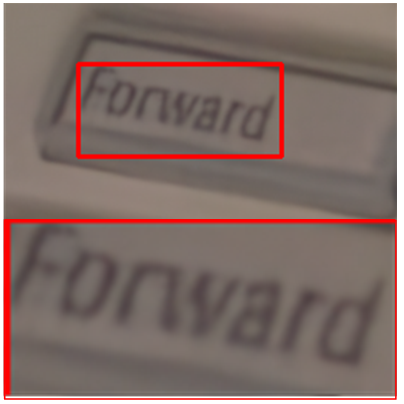}
    \includegraphics[width=1\linewidth]{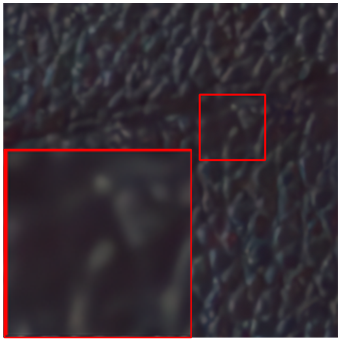}
    \includegraphics[width=1\linewidth]{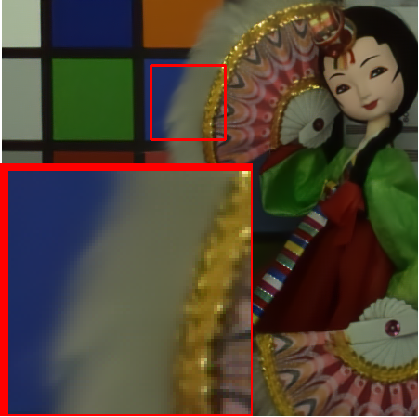}
    \end{minipage}}	
	\subfigure[InvDN]{\centering
	\begin{minipage}[b]{0.16\textwidth}
    \includegraphics[width=1\linewidth]{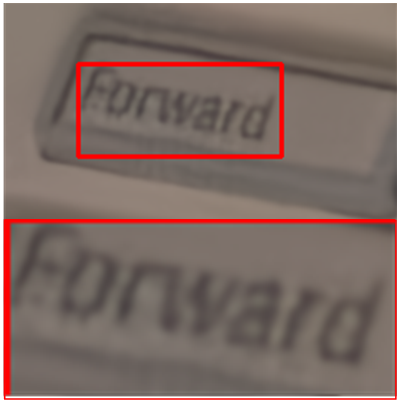}
    \includegraphics[width=1\linewidth]{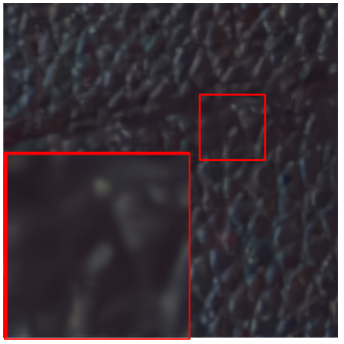}
    \includegraphics[width=1\linewidth]{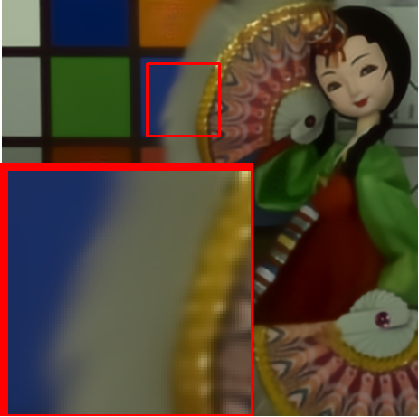}
    \end{minipage}}	
	\subfigure[\bf{Ours}]{\centering
	\begin{minipage}[b]{0.16\textwidth}
    \includegraphics[width=1\linewidth]{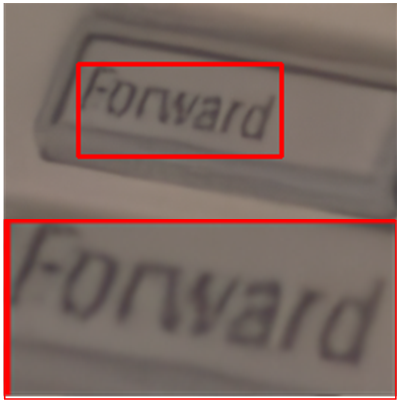}
    \includegraphics[width=1\linewidth]{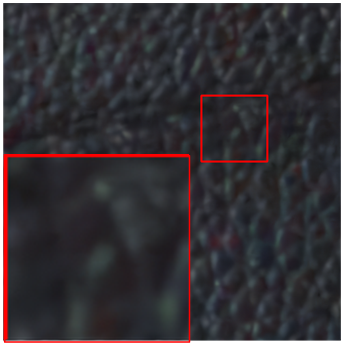}
    \includegraphics[width=1\linewidth]{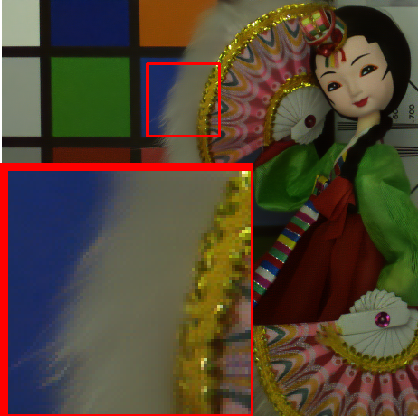}
    \end{minipage}}	
	\caption{Visual comparison of the results from real datasets. The first, second and third rows are the results from SIDD, DND and RNI15 datasets, respectively. The (a) noisy image is denoised by (b) CBDNet, (c) RIDNet, (d) VDN, (e) InvDN and (f) Our CFNet.}
	\label{fig5_SIDD}
\end{figure*}

\subsection{Result for Real Noise Removal}
In addition to the synthetic AWGN noise, real noise is always signal-dependent and spatially variant according to different ISP pipelines, which is more complicated. To further evaluate CFNet for real noise removal, we conduct the experiment on three real datasets, \textit{i.e.,} SIDD~\cite{sidd}, DND~\cite{dnd} and RNI15~\cite{rni}, which is compared with 3 traditional methods, \textit{i.e.,} BM3D~\cite{bm3d}, KSVD~\cite{ksvd}, WNNM~\cite{wnnm} and 17 CNN-based methods, \textit{i.e.,} DnCNN~\cite{dncnn}, CBDNet~\cite{guo2019toward}, RIDNet~\cite{ridnet}, VDN~\cite{vdn}, GradNet~\cite{gradnet}, DualBDNet~\cite{du2020blind}, AINDNet~\cite{aindnet}, DANet~\cite{danet}, InvDN~\cite{invdn},HI-GAN~\cite{higan}, GreneNet~\cite{grenenet}, DeamNet~\cite{deamnet}, DAGL~\cite{dagl}, MPRNet~\cite{mprnet}, MIRNet~\cite{zamirmirnet}, DBDNet~\cite{ma2021dbdnet} and CPNet~\cite{li2021cross}. To supervise the noise estimation, our model is first trained on both of synthetic data and real data (CFNet (All)), which is fine tuned on real SIDD dataset (CFNet (All-R)). For a fair comparison, we also list the results of the same kind of methods which explicitly estimates noise prior with All and All-R training schemes, \textit{e.g.,} CBDNet and AINDNet. The average PSNR and SSIM on SIDD and DND datasets are listed in Tab.~\ref{table_sidd} and~\ref{table_dnd}, respectively. All the results are referred from the original papers, except for CBDNet (All-R), AINDNet (All) and AINDNet (All-R) which are retrained by the code provided by the authors. Due to directly fine tuning on SIDD dataset, the models (All-R) reaches better results on SIDD dataset than the counterparts (All). 

\begin{figure}[t]
    \centering
    \includegraphics[width=0.94\columnwidth]{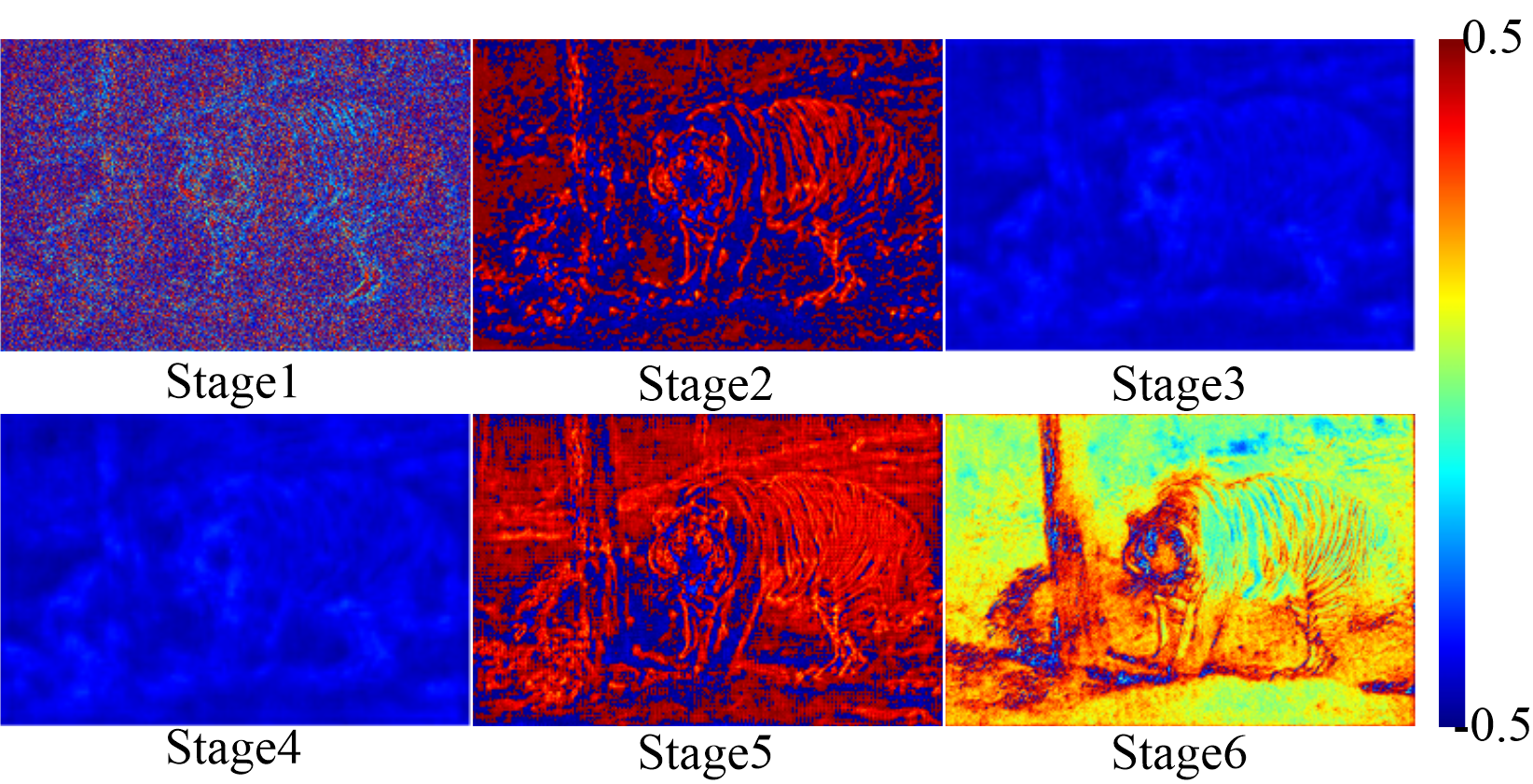}
    \caption{Visualization of the noise maps along Unet stages}
    \label{fig_Visualize the noise maps at stages}
\end{figure}

Overall, for SIDD dataset, GreneNet reaches the best result, and our CFNet and DeamNet show the comparable performance to GreneNet. However, as shown in the above experiment of synthetic noise removal, the methods which directly train the model on real SIDD dataset may face the risk of overfitting to noise prior of certain dataset. The significant improvement of our CFNet from 39.63 dB to 39.92 dB on DND dataset supports this observation. Also, compared with AINDNet whose performance is the best in the same kind of methods to ours, our CFNet provides the gaps of 0.24 dB and 0.44 dB for SIDD and DND datasets, respectively. Furthermore, the visual comparison is illustrated in Fig.~\ref{fig5_SIDD}. There is significant noise observed on the results of CBDNet. By contrast, our results show the superior performance for noise removal, detail recovery and edge preserving than CBDNet, RIDNet, VDN and InvDN.

\section{Discussion}
To further comprehend the insights of the proposed CFNet, this section first provides the visualization comprehension. Then, the hyper-parameter study and ablation study are conducted. Finally, the complexity is compared with SOTAs.

\begin{figure}[t]
    \centering
    \includegraphics[width=0.94\columnwidth]{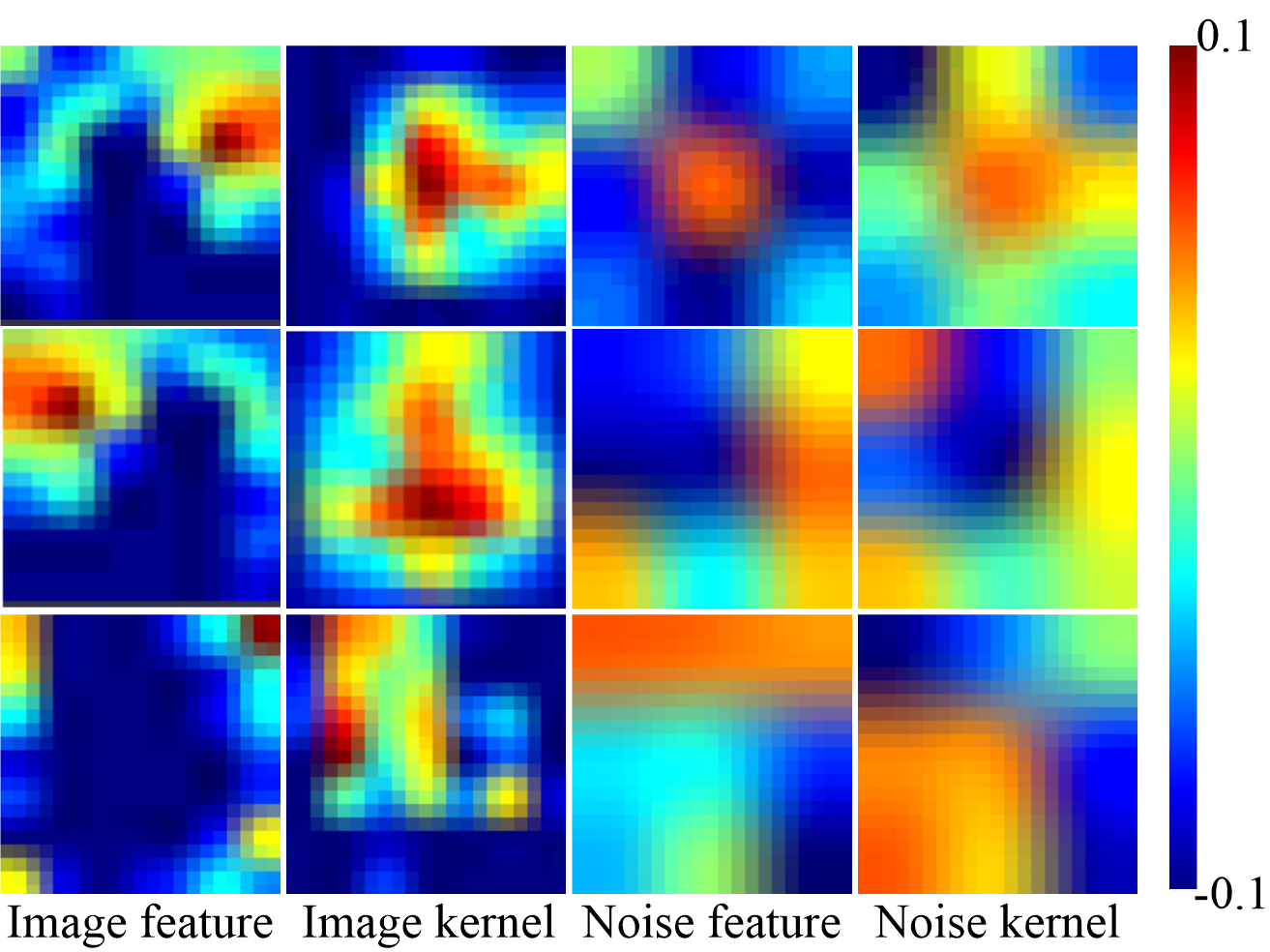}
    \caption{Visualization of conditional kernels}
    \label{fig_Visualize the kernel}
\end{figure}

\subsection{Visualization Comprehension}
The noise features along Unet stages and conditional kernels are visualized to validate the effectiveness of dynamic noise estimation and adaptive denoising, respectively.

\subsubsection{Noise Features at Stages}
Since only the first noise estimation module has supervision, Fig.~\ref{fig_Visualize the noise maps at stages} shows the samples of the corresponding noise features via heat maps instead of the estimated noise maps. Along the stages of Unet, the messy noise features are firstly encoded in high-dimension space, and then followed by a smooth representation. So, the noise can be progressively mitigated.

\begin{table*}[t]\footnotesize 
	\centering
	\caption{The study of the hyper-parameters $t$, $g$ and $k$. All the variants are trained in the same way as CFNet (All) and tested on SIDD dataset via online submission}
	\label{tab:study_trk}
	\setlength{\tabcolsep}{2mm}{
	\begin{tabular}{c|c|c|c|c|c|c|c|c|c}
	    \hline
		\multirow{2}*{Metrics} & \multicolumn{3}{c|}{$(t, g=16, k=3)$} &
		\multicolumn{3}{c|}{$(g, t=2, k=3)$} & \multicolumn{3}{c}{$(k, t=2, g=16)$} \\ 
		\cline{2-10}
		& 1 & 2 & 3 & 8 & 16 & 32 & 3 & 5 & 7 \\
	    \hline 
		PSNR & 38.95 & \bf{39.09} & 39.09 & 39.02 & \bf{39.09} & 38.92 & \bf{39.09} & 39.07 & 39.06 \\
		\hline
		SSIM  & 0.952 & \bf{0.953} & 0.953 & 0.953 & \bf{0.953} & 0.951 & \bf{0.953} & 0.953 & 0.953 \\
		\hline
	\end{tabular}}
\end{table*}

\subsubsection{Conditional Kernels}
We visualize the samples of image feature, noise feature and their conditional kernels in Fig.~\ref{fig_Visualize the kernel}. For better illustration, the original 3$\times$3 resolution is interpolated via bicubic to 21$\times$21. Since residual learning is introduced in conditional filter block, all the conditional kernels follow the style of high-pass filters. Specially, it is shown that the learned image kernels are edge-preserving, and the corresponding noise kernels are adaptive to noise prior. Thus, the proposed conditional kernels can efficiently infer the content-aware kernels for all local features.

\subsection{Hyper-parameters Study}
This part studies the influence of the key hyper-parameters. it includes number of conditional filter blocks $t$ within each conditional denoising module, channel group number of conditional denoising kernels $g$ and the learned conditional kernel size $k$. The hyper-parameters are searched by fixing the others. The corresponding models are trained in the same way as CFNet (All), and evaluated on SIDD dataset~\cite{sidd} shown in Tab.~\ref{tab:study_trk}. The optimal channel group number of conditional denoising kernels $g$ is 16. Also, it is observed that there is no significant gain as increasing number of conditional filter blocks $t$ larger than 2. In addition, our CFNet is insensitive to the conditional kernel size $k$. Therefore, to balance performance and complexity, we choose $\left(t=2, g=16, k=3\right)$ in the proposed CFNet.

\begin{table}[t]\footnotesize
\setlength{\abovecaptionskip}{0pt}
\centering 
\caption{Ablation study on the variants of our CFNet with ($\checkmark$) or without ($\xmark$) the key contributions. All the variants are trained in the same way as CFNet (All) and tested on SIDD dataset via online submission.\\[3pt]}
\label{table_AblationStudy}
\setlength{\tabcolsep}{0.6mm}{
\begin{tabular}{c|ccccccccc}
\hline
ATBs&\checkmark&\checkmark&\xmark&\checkmark&\checkmark&\xmark&\xmark&\xmark&\\
CFBs&\checkmark&\checkmark&\checkmark&\xmark&\xmark&\checkmark&\xmark&\xmark&\\
DNE&\checkmark&\xmark&\checkmark&\checkmark&\xmark&\xmark&\checkmark&\xmark&\\
\hline
PSNR&\bf{39.09}&38.86&39.00&38.57&38.47&38.47&38.47&38.30\\
\hline
\end{tabular}}
\end{table}

\begin{table}[t]\footnotesize  
	\centering
	\caption{Ablation study on the variants of NEMs. All the variants are trained in the same way as CFNet (All) and tested on SIDD dataset via online submission}  
	\label{tab:ATB_NEM_sharing.}
    \setlength{\tabcolsep}{3mm}{
	\begin{tabular}{c|c|c|c|c|c}
		\hline
		\multirow{2}*{Metrics}& \multicolumn{3}{c|}{Number of NEMs} &\multicolumn{2}{c}{Sharing Setting}  \\
		\cline{2-6}
		&1&3&6&sharing&no sharing \\
		\hline
		PSNR& 38.86 & 38.94 & \bf{39.09} & \bf{39.09} & 38.93\\
		\hline
	\end{tabular}}
\end{table}

\begin{figure}[t]
    \centering
    \includegraphics[width=1\columnwidth]{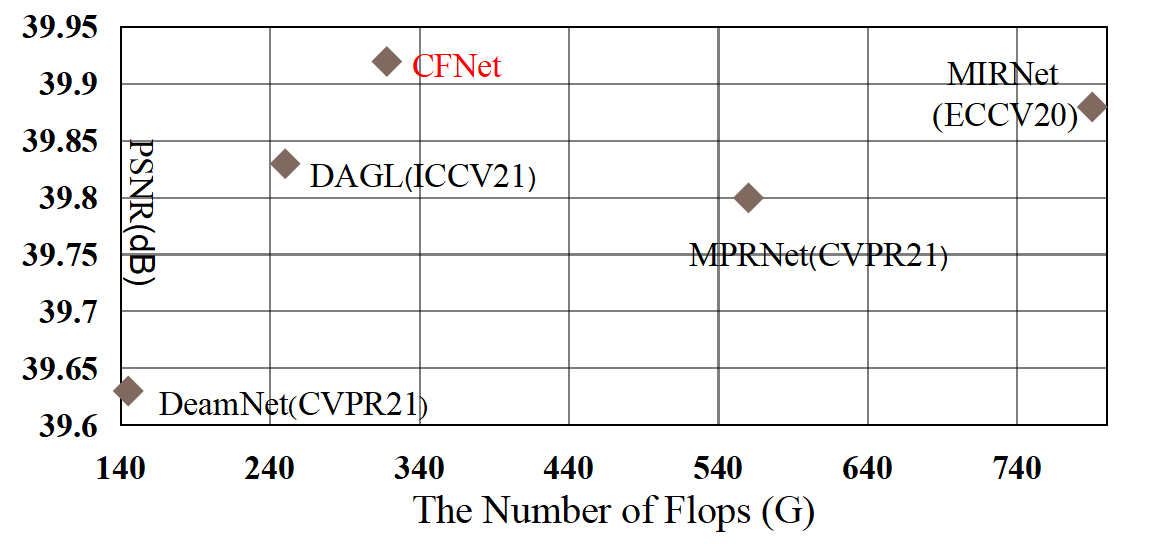}
    \caption{Comparison of GFLOPs and PSNR on DND dataset}
    \label{CFNet_flops}
\end{figure}

\subsection{Ablation Study}
\label{sec:Ablation Study}
The main contributions are related to Affine Transform Blocks (ATBs) for noise estimation, Conditional Filter Blocks (CFBs) for adaptive denoising and stage-wise estimation modules for Dynamic Noise Estimation (DNE). In this part, we further validate their effectiveness to the proposed CFNet. For fair comparison, the baseline uses ten convolutional layers, one-off noise estimation and a convolutional layer on the concatenated features to replace each ATB, DNE and CFB, respectively. All the models are trained in the same way as CFNet (All). The performance of the corresponding variants by combining different contributions is listed in Tab.~\ref{table_AblationStudy}. It is observed that the baseline without any contribution shows the worst result. The improvement is progressively achieved as more contribution is introduced. However, some interesting phenomenons should be noted to understand the roles of all contributions. Since the noise estimation is inaccurate without dynamically updating, the effect of CFBs is suppressed. In addition, due to the unreasonable noise feature concatenating, although the network is guided by accurate and dynamic noise guidance, the result is still sub-optimal. Thanks to accuracy of dynamic noise estimation and power of conditional denoising, the significant improvement of our CFNet is observed. 

In addition, to explore the effect of update frequency on noise prior, Tab.~\ref{tab:ATB_NEM_sharing.} compares static NEM, semi-dynamic NEM and dynamic NEM for the first stage, the former three stages and all the stages, respectively. It is shown that our dynamic NEM reaches the best performance. Also, by learning noise estimation in the first NEM, the following NEMs update noise prior via parameter sharing, which is not only reasonable, but also efficient as shown in Tab.~\ref{tab:ATB_NEM_sharing.}.

\subsection{Complexity Comparison}
Fig.~\ref{CFNet_flops} shows GFLOPs and PSNR of the proposed CFNet on DND dataset, which compares with 4 SOTAs, \textit{i.e.,} DeamNet~\cite{deamnet}, DAGL~\cite{dagl}, MPRNet~\cite{mprnet} and MIRNet~\cite{zamirmirnet}. Specially, MPRNet shows inferior performance to our CFNet with significant higher GFLOPs, and MIRNet reaches comparable performance to ours at the cost of extra 490 GFLOPs. DeamNet and DAGL which have lower GFLOPs with observed performance degradation. Thus, it indicates that our method reaches a good balance between efficiency and effect.

\section{Conclusion}
This paper proposes CFNet with three contributions. First, based on features from the image and the noise map, a novel conditional filter block is designed to adaptively infer the denoising kernels for local windows centered by all feature positions. In addition, in the CNN structure, noise estimation and non-blind denoising are alternatively performed to continuously update noise prior as iterative feature denoising. Furthermore, iterative affine transform blocks are proposed to predict synthetic heteroscedastic Gaussian noise distribution. Based on qualitative and quantitative evaluations, comprehensive experiments on mainstream synthetic datasets and real datasets show improvement of the proposed CFNet when compared with SOTAs.

Very recently, diffusion model is quite hot in low-level vision which is more robust than GAN. In the future, we will investigate the ways for image denoising in the framework of conditional diffusion model. The key problems are two-folds. One is to tackle the noise whose distribution is non-Guassian, and the other is to further accelerate the inference time of diffusion model. 





\bibliographystyle{elsarticle-num}
\bibliography{egbib}

\vfill
\end{document}